\documentclass[useAMS,usenatbib,fleqn]{mn2e}  
\usepackage{times}
\usepackage{graphicx}
\usepackage{amsmath}
\def\bt{{\mbox{\boldmath $\theta$}}}

\def\bx{{\mbox{\boldmath $x$}}}
\def\xmm{{\it XMM-Newton}}
\def\E{\mathrm{E}}
\def\V{\mathrm{V}}
\def\gsim{\mathrel{\hbox{\rlap{\hbox{\lower4pt\hbox{$\sim$}}}\hbox{$>$}}}}
\def\lsim{\mathrel{\hbox{\rlap{\hbox{\lower4pt\hbox{$\sim$}}}\hbox{$<$}}}}
   \title[A Bayesian test for periodic signals in red noise]{A Bayesian test for periodic signals in red noise}
   \author[S. Vaughan]{S. Vaughan\thanks{E-mail: sav2@star.le.ac.uk}\\
              X-Ray and Observational Astronomy Group, University of
              Leicester,  Leicester, LE1 7RH, U.K.
             }

\begin{document}

 \date{Accepted 2009 October 14. Received 2009 October 14; in original form 2009 September 11}
 
\pagerange{\pageref{firstpage}--\pageref{lastpage}} \pubyear{2009}

\maketitle

\label{firstpage}

\begin{abstract}
   Many astrophysical sources, especially compact accreting sources, 
   show strong, random brightness fluctuations with broad power spectra in addition to periodic 
   or quasi-periodic oscillations (QPOs) that have narrower spectra. The random nature of the dominant source 
   of variance greatly complicates the process of searching for possible weak periodic signals.
   We have addressed this problem using the tools of Bayesian statistics; in particular
   using Markov chain Monte Carlo techniques to approximate the posterior distribution of model
   parameters, and posterior predictive model checking to assess model fits and search 
   for periodogram outliers that may represent periodic signals.
   The methods developed are applied to two example datasets, both long \xmm\ observations of
   highly variable Seyfert 1 galaxies: RE J$1034+396$ and Mrk $766$. In both cases a
   bend (or break) in the power spectrum is evident. In the case of RE J$1034+396$ the
   previously reported QPO is found but with somewhat weaker statistical significance
   than reported in previous analyses. The difference is due partly to the improved continuum
   modelling, better treatment of nuisance parameters, and partly to different data selection methods.
\end{abstract}

\begin{keywords}
   Methods: statistical -- Methods: data analysis  -- X-rays: general -- Galaxies: Seyfert
\end{keywords}

   \maketitle
%
%________________________________________________________________

\section{Introduction}
\label{sect:intro}

\defcitealias{Vaughan05}{V05}

A perennial problem in observational astrophysics is detecting
periodic or almost-periodic signals in noisy time series. 
The standard analysis tool is the periodogram \citep[see e.g. ][]{Jenkins69, Priestley81,
  Press92, Bloomfield00, Chatfield03}, and the problem of
period detection
amounts to assessing whether or not some particular peak in the
periodogram is due to a periodic component or a random fluctuation in
the noise spectrum \citep[see][]{Fisher29, Priestley81, Leahy83,
  vanderklis89, Percival93, Bloomfield00}. 

If the time series is the sum of a random
(stochastic) component and a periodic one we may write $y(t) = y_R(t)
+ y_P(t)$ and, due to the independence of $y_R(t)$ and $y_P(t)$, the
power spectrum of $y(t)$ is the sum of the two power spectra of the
random and stochastic processes: $S_Y(f) = S_R(f) + S_P(f)$. 
This is a \emph{mixed} spectrum \citep[][section 4.4]{Percival93} formed from the sum of 
$S_P(f)$, which comprises only narrow features, and 
$S_R(f)$, which is a continuous, broad spectral function.
Likewise, we may consider an
evenly sampled, finite time series $y(t_i)$ ($i=1,2,\ldots,N$) 
as the sum of two finite time series: one is a realisation of the
periodic process, the other a random realisation of the stochastic process.
We may
compute the periodogram (which is an estimator of the true power
spectrum) from the squared modulus of the Discrete Fourier Transform
(DFT) of the time series, and, as with the power spectra, the periodograms of the
two processes add linearly: $I(f_j) = I_R(f_j) + I_P(f_j)$. The periodogram of the periodic time series
will contain only narrow ``lines'' with all the power concentrated in
only a few frequencies, whereas the periodogram of the
stochastic time series will show power spread over many frequencies. 
Unfortunately the periodogram of stochastic processes 
fluctuates wildly around the true power spectrum, making it
difficult to distinguish
random fluctuations in the noise spectrum from truly spectral periodic
components. See \cite{vanderklis89} for a thorough review of these
issues in the context of X-ray astronomy.

Particular attention has been given to the special case that the spectrum
of the stochastic process is 
flat (a \emph{white noise} spectrum $S(f) = const$), which is the case
when the time series data $y_R(t_i)$
are independently and identically distributed (IID) random variables.
Reasonably well-established statistical procedures have been developed
to help identify spurious spectral peaks and reduce the chance of
false detections \citep[e.g.][]{Fisher29, Priestley81, Leahy83,
  vanderklis89, Percival93}. 
In contrast there is no comparably well-established procedure in the general case
that the spectrum
of the stochastic process is not flat. 

In a previous paper,
\cite{Vaughan05} (henceforth \citetalias{Vaughan05}), we proposed what is essentially a
generalisation of Fisher's method to the case where the noise spectrum
is a power law: $S_R(f) = \beta f^{-\alpha}$ (where $\alpha$ and
$\beta$ are the power law index and normalisation parameters). 
Processes with power spectra that show a power law dependence on frequency with
$\alpha > 0$ (i.e. increasing power to lower frequencies) are called
\emph{red noise} and are extremely common in astronomy and elsewhere
\citep[see][]{Press78}. In this paper we expand upon the ideas in
\citetalias{Vaughan05} and, in particular, address the problem from a
Bayesian perspective that allows further generalisation of the
spectral model of the noise.

The rest of this paper is organised as follows.
In section~\ref{sect:bayes} we introduce some of the basic concepts of 
the Bayesian approach to statistical inference; readers familiar with
this topic may prefer to skip this section.
Section~\ref{sect:stat} gives a brief overview of classical significance
testing using $p$-values (tail area probabilities) and test statistics, and 
section~\ref{sect:ppp} discusses the posterior predictive $p$-value, a Bayesian
counterpart to the classical $p$-value.
Section \ref{sect:pc} reviews the conventional (classical)
approaches to testing for periodogram peaks.
Section \ref{sect:ml} outlines the theory of maximum likelihood estimation
from periodogram data, which is developed into the basis of a fully Bayesian analysis in sections~\ref{sect:ba} and \ref{sect:ppper}. 
The Bayesian method is then applied to two real observations if AGN in section \ref{sect:data}.
Section \ref{sect:disco} discusses the limitations of the method, and alternative approaches
to practical data analysis. A few conclusions are given in section \ref{sect:conc}, and 
 two appendices describe details of the simulations algorithms used
in the analysis.

%__________________________________________________________________

\section{Bayesian basics, briefly}
\label{sect:bayes}

\begin{table*}
  \caption[]{Definitions used throughout the paper.}
  \label{table:def}
  \centering
  \begin{tabular}{l l}
  \hline
  \hline
  Term & Definition \\
  \hline
  $f_j$   & The $j$th Fourier frequency $f_j = j / N \Delta T$ ($j=1,\ldots,N/2$)\\
  $I_j$   & Periodogram at frequency $f_j$ \\
  $\mathbf{I}$ & vector of periodogram values $\mathbf{I}= \{ I_1,\ldots,I_{N/2} \}$ \\
  $\bt$   & Model parameters $\bt = \{ \theta_1, \ldots, \theta_M \}$ \\
  $\hat{\bt}_{\rm MLE}$   & Maximum Likelihood Estimates of parameters (equation \ref{eqn:mle}) \\
  $\bx$   & Data (e.g. time series) $\bx = \{ x_1, \ldots, x_N \}$ \\
  $p_C$   & Frequentist/classical (conditional) $p$-value (equation \ref{eqn:classic})\\
  $p_B$   & Bayesian (posterior predictive) $p$-value (equation \ref{eqn:ppp})\\
  $S_j$   & Model spectral density at frequency $f_j$, i.e. $S(f_j; \bt)$ \\
  $\hat{S}_j$ & The model computed at the estimate $\hat{\bt}$ (equation \ref{eqn:Rstat})\\
  $D(\bx, \bt)$ & Deviance ($-2 \log p(\bx | \bt)$) given model $H$ (equation \ref{eqn:mlogl}) \\
  $p(\bx | \bt, H)$ & Likelihood for parameters $\bt$ of model $H$ given data $\bx$ (equation \ref{eqn:bayes_eqn}) \\
  $p(\bt|H)$ & Prior probability density for parameters $\bt$ (equation \ref{eqn:bayes_eqn}) \\
  $p(\bt | \bx, H)$ & Posterior probability density for parameters $\bt$ given data $\bx$ (equation \ref{eqn:bayes_eqn}) \\
  $p(\bx | H)$ & Prior predictive density (aka marginal likelihood) of the data $\bx$ (equation \ref{eqn:bayes_eqn}) \\
  $\bx^{\rm rep}$ & Replicated data (from repeat observations or simulations) (equation \ref{eqn:ppd})\\
  $p(\bx^{\rm rep} | \bx^{\rm obs})$ & Posterior predictive distribution given data $\bx^{\rm obs}$ (equation \ref{eqn:ppd})\\
  $T(\mathbf{x})$ & A test statistic \\
  \hline
  \end{tabular}
\end{table*}

There are two main tasks in statistical inference: parameter
estimation and model checking (or comparison). 
Bayesian parameter estimation is concerned with finding the probability of
the parameters given the model $p(\bt | \bx, H)$, where $\bx$ ($= \{ x_1, \ldots, x_N \}$) are 
data values , $\bt$ ($= \{ \theta_1, \ldots, \theta_M \}$) are parameter
values  and $H$ represents the model. In contrast,
\emph{frequentist} (or \emph{classical}) statistics restricts
attention to the sampling distribution of the data given the model and
parameters $p(\bx | \bt, H)$.
These two probability functions are related by Bayes' Theorem 
\begin{equation}
p(\bt | \bx, H) = \frac{p( \bx | \bt, H) p( \bt| H)}{p(\bx | H)}.
\label{eqn:bayes_eqn}
\end{equation}
Each of the terms in Bayes' theorem has a name when used in Bayesian
data analysis: $p(\bt | \bx, H)$ is the \emph{posterior} distribution of the
parameters; $p( \bx | \bt, H)$ is the \emph{likelihood} function of the
parameters\footnote{
Note that when considered as a function of the data for known
parameters, $p( \bx | \bt, H)$
is the sampling distribution of the data, but when considered as a function of the
parameters for fixed data, $p( \bx | \bt, H)$ is known as the
likelihood, sometimes denoted $L(\bt)$.}; 
$p( \bt| H)$ is the \emph{prior} distribution of the parameters,
and $p(\bx | H)$ is a normalising constant sometimes referred to as
the \emph{marginal likelihood} (of the data) or 
the \emph{prior predictive} distribution\footnote{Some physicists use the \emph{evidence} for this term, e.g. \citet{Sivia96}, \citet{Trotta08}.}.
General introductions to Bayesian analysis for the non-specialist include
\citet{Jeffreys92}, \citet{Berger88} and \citet{Howson91}; more
thorough treatments include \cite{Berry96}, \citet{Carlin00},
\citet{Gelman04}, and \citet{Lee89}; and discussions more focussed on physics and
astrophysics problems include \citet{Sivia96}, \citet{Gregory05}, and \citet{Loredo90,
  Loredo92}. 

In Bayesian analysis the posterior distribution is a complete summary of our
inference about the parameters given the data $\bx$, model
$H$, and any prior information.
But this can be further summarised using a point estimate 
for the parameters such as the mean, median or mode of the posterior distribution.
For one parameter the posterior mean is
\begin{equation}
\label{eqn:post-mean}
  \E[ \theta | \bx, H ]  = \int \theta p(\theta | \bx, H) d\theta
\end{equation}
A slightly more informative summary is a \emph{credible interval} 
(or credible region for multiple parameters). This is an interval in parameter space that contain a specified 
probability mass (e.g. $90$ per cent) of the posterior distribution. These intervals give an indication of the uncertainty 
on the point inferences.
\begin{equation}
\label{eqn:ci}
  \int_R p( \theta | \bx, H) d\theta = C,
\end{equation}
where $C$ is the probability content (e.g. $C=0.9$) and $R$ is interval in parameter space.
One common approach is to select the interval satisfying equation~\ref{eqn:ci} that contains the 
highest (posterior) density (i.e. the posterior density at any point inside is higher than at any point outside). 
This will give the smallest interval that contains a probability $C$, usually called the highest posterior density region 
(abbreviated to HDR or HPD interval by different authors). An alternative is the equal tail posterior interval, 
which is defined by the two values above and below which is $(1-C)/2$ of the posterior probability. These two types of
interval are illustrated in \citet[][see their Fig. 1]{Park08}.

If we have multiple parameters but are interested in only one parameter we may \emph{marginalize} over the other parameters. For 
example, if $\bt = \{ \theta_1, \theta_2 \}$ then the posterior distribution for $\theta_1$ is
\begin{eqnarray}
\label{eqn:margin}
  p( \theta_1 | \bx , H ) & = & \int p( \theta_1, \theta_2 | \bx, H ) d\theta_2 \nonumber \\
                          & = & \int p( \theta_1 | \theta_2, \bx, H ) p(\theta_2 | \bx, H) d\theta_2.
\end{eqnarray}
This is the average of the \emph{joint} posterior $p( \theta_1, \theta_2 | \bx, H )$ over $\theta_2$.
In the second formulation the joint posterior has been factored into two distributions, the first is the 
conditional posterior of $\theta_1$ given $\theta_2$ and the second is the posterior density for $\theta_2$.

Most present day Bayesian analysis is carried out with the aid of Monte Carlo methods for evaluating the necessary integrals.
In particular, if we have a method for simulating a random sample of size $N$ from the posterior distribution $p(\theta | \bx, H)$ then 
the posterior density may be approximated by a histogram of the random draws. 
This gives essentially complete information about the posterior (for a sufficiently large $N$). 
The posterior mean may be 
approximated by the sample mean
\begin{equation}
\E[\theta | \bx , H] \approx \frac{1}{N} \sum_{i=1}^{N} \theta^i
\end{equation} 
where $\theta^i$ are the individual simulations from the posterior. If the parameter is a vector $\bt = \{ \theta_1, \ldots, \theta_M \}$, 
the $m$th component of each vector is a sample from the marginal distribution of the $m$th parameter. This means the 
posterior mean of the each parameter is approximated by the sample mean of each component of the vector.
Intervals may be calculated from the sample quantiles, e.g. the $90$ per cent equal tail area interval on a parameter may be
approximated by the interval between the $0.05$ and $0.95$ quantiles of the sample.
In this manner the difficult (sometimes insoluble) integrals of equations \ref{eqn:post-mean}, \ref{eqn:ci} and \ref{eqn:margin} may be replaced
by trivial operations on the random sample. The accuracy of these approximations is governed by the accuracy with which
the distribution of the simulations matches the posterior density, and the size of the random sample $N$. 
Much of the work on practical Bayesian data analysis methods has been devoted to the generation and assessment of accurate
Monte Carlo methods, particularly the use of Markov chain Monte Carlo  (MCMC) methods, which will be discussed and used later in this paper.

For model comparison we may again use Bayes theorem to give the  
posterior probability for model $H_i$
\begin{equation}
p(H_i | \bx) = \frac{ p( \bx | H_i) p(H_i) }{ p(\bx) },
\end{equation}
and then compare the posterior probabilities for two
(or more) competing models, say $H_0$ and $H_1$ (with parameters
$\bt_0$ and $\bt_1$, respectively). 
(In effect we are treating the choice of model, $H_i$, as a discrete parameter.)
The ratio of these two eliminates the
term in the denominator (which has no dependence on model selection):
\begin{equation}
O = \frac{ p(H_1 | \bx )}{ p(H_0 | \bx )} 
  =  \frac{p(\bx | H_1)}{ p(\bx | H_0)}  \frac{p(H_1)}{p(H_0)}
\label{eqn:oddsratio}
\end{equation}
The first term on the right hand side of
equation~\ref{eqn:oddsratio} is the ratio of likelihoods and is
often called the \emph{Bayes factor} \citep[see][]{Kass95} and the second term is the ratio of the
priors.
However, in order to obtain $p(H_i|\bx)$ we must first remove the dependence of the posterior
distributions on their parameters, often called \emph{nuisance} parameters in
this context (we are not interested in making inferences about $\bt_i$, but they
are necessary in order to compute the model).
In order to do this the full likelihood function
must be integrated or {\it marginalized} over the joint prior probability
density function (PDF) of the parameters:
\begin{equation}
p(\bx | H_i) = \int p( \bx | \bt_i, H_i) p(\bt_i | H_i) d\bt_i
\label{eqn:marginal}
\end{equation}
Here, $p(\bx | \bt_i, H_i)$ is the likelihood 
and $p(\bt_i | H_i)$ the prior
for the parameters of model $H_i$.

%__________________________________________________________________

\section{Test statistics and significance testing}
\label{sect:stat}

We return briefly to the realm of frequentist statistics and consider
the idea of significance testing using a test statistic.
A test statistic $T(\bx)$ is a real-valued function of the data
chosen such that extreme values are unlikely when the \emph{null
  hypothesis} $H_0$ is true. If the sampling distribution of $T$ is
$p(T|H_0)$, under the null hypothesis, and the observed value is $T^{\rm obs} = T(\bx^{\rm obs})$, then the 
classical $p$-value is 
\begin{equation}
p_C(\bx^{\rm obs}) = \int_{T^{\rm obs}}^{+\infty} p(T|H_0) dT 
    = \Pr \{ T(\bx^{\rm rep}) \ge T(\bx^{\rm obs}) | H_0 \},
\end{equation}
where $\Pr{x|y}$ is the probability of event $x$ given that event $y$ occured.
The second formulation is in terms of \emph{replicated} data that
could have been observed, or could be observed in repeat experiments
\citep{Meng94, Gelman96, Gelman04}.
The $p$-value gives the fraction of $p(T|H_0)$ lying above the observed
value $T^{\rm obs}$. As such, $p$-values are \emph{tail area}
probabilities, and 
one usually uses small $p_C$ as evidence against the null hypothesis.
If the null hypothesis is \emph{simple}, i.e. has no free parameters, or the sampling
distribution of $T$ is independent of any free parameters, then the
test statistic is said to be \emph{pivotal}. If the distribution of
the test statistic does depend on the parameters of the model, i.e $p(T| \bt, H_0)$, as is
often the case, then we have a \emph{conditional} $p$-value
 \begin{equation}
\label{eqn:classic}
p_C(\bx^{\rm obs}, \bt) = \int_{T^{\rm obs}}^{+\infty} p(T|\bt) dT
         = \Pr \{ T(\bx^{\rm rep}) \ge T(\bx^{\rm obs}) | \bt \}
\end{equation}
(For clarity we have omitted the explicit conditioning on $H_0$.)
In order to compute this we must have an estimate for the nuisance parameters
$\bt$.

%__________________________________________________________________

\section{Posterior predictive $p$-values}
\label{sect:ppp}

In Bayesian analysis the \emph{posterior predictive distribution} is
the distribution of $\bx^{\rm rep}$ given the available information
which includes $\bx^{\rm obs}$ and any prior information.
\begin{equation}
\label{eqn:ppd}
p( \bx^{\rm rep} | \bx^{\rm obs}) = \int p( \bx^{\rm rep} | \bt ) p(
\bt | \bx^{\rm obs}) d\bt,
\end{equation}
\citep[e.g. section 6.3 of][]{Gelman04}. 
Here, $p(\bt | \bx^{\rm obs})$ is the posterior distribution of the
parameters (see eqn.~\ref{eqn:bayes_eqn}) and $p( \bx^{\rm rep} |
\bt )$ is the sampling distribution of the data given the parameters. 
The Bayesian $p$-value is the (tail area) probability that replicated
data could give a test statistic at least as extreme as that observed.
\begin{eqnarray}
\label{eqn:ppp}
p_B(\bx) & = & \int p_C(\bx^{\rm obs}, \bt) p(\bt | \bx^{\rm obs}) d\bt \nonumber \\
    & = & \Pr \{ T(\bx^{\rm rep}) \ge T(\bx^{\rm obs}) | \bx^{\rm obs}, H_0 \}
\end{eqnarray}
This is just the classical $p$-value (eqn.~\ref{eqn:classic}) averaged
over the posterior distribution of $\bt$
(eqn.~\ref{eqn:bayes_eqn}), i.e. the posterior mean $\E[p_C|\bx^{\rm obs}, H_0]$
which may be calculated using simulations.
In other words, it gives the average of the conditional $p$-values evaluated
at over the range of parameter values, weighted by the (posterior) probability
of the parameter values.
The aim of the posterior predictive $p$-value (or, more generally, comparing the
observed value of a test statistic to its posterior
predictive distribution) is to provide a simple assessment of whether the 
data are similar (in important ways) to the data expected under a particular model.

This tail area probability does not depend on the unknown value of
parameters $\bt$, and is often called the \emph{posterior predictive}
$p$-value \citep[see][]{Rubin84, Meng94, Gelman96, Gelman04, Protassov02}.
The (classical) conditional $p$-value and the (Bayesian) posterior
predictive $p$-value are in general different but are equivalent 
in two special cases.
If the null hypothesis is simple or the test statistic $T$ is pivotal, 
then the sampling and posterior predictive
distributions of $T$ are the same, $p_C = p_B$.

Like the classical (conditional) $p$-value, $p_B$ is used for model
checking but has the advantage of having no dependence on unknown
parameters. The posterior predictive distribution of $T$
includes the uncertainty in the classical $p$-value $p_C$
due to the unknown nuisance parameters \citep{Meng94}.

The posterior predictive $p$-value is a single summary of the 
agreement between data and model, and may be used to assess 
whether the data are consistent with being  drawn from the 
model: a $p$-value that is not extreme (i.e. not very close to $0$ or $1$) 
shows the observed value $T^{\rm obs}$ is not an outlier in the 
population $T^{\rm rep}$. \citet{Gelman04} and \citet{Protassov02} argue that model
checking based on the posterior predictive distribution is less
sensitive to the choice of priors (on the parameters), and more useful
in identifying deficiencies in the model, compared to Bayes factors or
posterior odds (eqn.~\ref{eqn:oddsratio}).

%__________________________________________________________________

\section{Conditional significance of periodogram peaks}
\label{sect:pc}

We now return to the problem of assessing the significance of peaks in
periodograms of noisy time series. The null hypothesis, $H_0$, in this 
case is that the time series was the product of a stochastic
process. It is well known that the periodogram of any stochastic time
series of length $N$, denoted $I_j = I(f_j)$ at 
Fourier frequency $f_j = j/N \Delta T$ (with $j=1,\ldots,N/2$), is 
exponentially distributed\footnote{
The exponential distribution $p(x|\lambda) = \lambda e^{-\lambda x}$ is 
a special case of the chi square distribution $\chi_{\nu}^2$ with $\nu=2$ 
degrees of freedom, and a special case of the gamma distribution, $\Gamma(1,1/\lambda)$.
See e.g. \citet{Eadie71}, \citet{Carlin00}, \citet{Gelman04} or \citet{Lee89}
for more on specific distribution functions.
}
about the true spectral density $S_j = S(f_j)$
\begin{equation}
\label{eqn:pdist}
  p(I_j | S_j) = \frac{1}{S_j} \exp( -I_j / S_j ),
\end{equation}
\citep[see][]{Jenkins69, Groth75, Priestley81, Leahy83, vanderklis89, Press92, Percival93,
  Timmer95, Bloomfield00, Chatfield03}.
Strictly speaking this is valid for the Fourier frequencies other than the
zero and Nyqist frequency ($j=1$ and $j=N/2$), which follow a different
distribution, although in the limit of large $N$ this difference is almost always inconsequential.
This distribution means that the ratio of the periodogram ordinates
$I_j$ to the true spectrum $S_j$ will be identically distributed.
If we have a parametric spectral model with known parameters,
  $S_j(\bt)$, the ratio
\begin{equation}
  R_j^{\rm obs} = 2 I_{j}^{\rm obs} / S_j(\bt)
\end{equation}
will be distributed as $\chi_{\nu}^2$ with $\nu = 2$ degrees of
freedom (see \citetalias{Vaughan05})
and it is trivial to integrate this density function to find the classical tail area
$p$-value corresponding to a given observed datum $I_j^{\rm obs}$
This simple fact is the basis of many ``textbook'' frequentist tests for periodicity.
However, $p_C$ depends the parameters $\bt$ (and, more generally, the model $H$),
which in general we do not know. 

The standard solution is to estimate the parameters, e.g. by fitting the periodogram data, 
and thereby estimate the spectral density $S_j$ under the null hypothesis, call this $\hat{S}_j$, and use this
estimate in the test statistic
\begin{equation}
\label{eqn:Rstat}
  \hat{R}_j^{\rm obs} = 2 I_{j}^{\rm obs} / \hat{S}_j.
\end{equation}
The problem is that the distribution of $\hat{R}_j$ will not be simply $\chi_2^2$
since that does not account for the uncertainty in the spectral estimate $\hat{S}_j$.
\citetalias{Vaughan05} presented a partial solution to this, by treating the statistic $\hat{R}_j$ as the ratio
of two random variables under certain simplifying assumptions. 
In what follows we use Bayesian methods to develop a much more general method for estimating the parameters of a power
spectral model, 
and posterior predictive model checking to check the
quality of a model fit and to map out the distribution of $\hat{R}_j$ conditional on the observed data.

%__________________________________________________________________

\section{Periodogram analysis via the likelihood function}
\label{sect:ml}

As discussed in \citetalias{Vaughan05}, and based on the results of \citet{Geweke83}, 
a very simple way to obtain a reasonable estimate of the index and normalisation
of a power law power spectrum, $S(f) = \beta f^{-\alpha}$, is by linear regression of $\log I_j^{\rm obs}$ on $\log f_j$ 
\citep[see also][]{Pilgram98}.
This provides approximately unbiased and normally distributed estimates of the power law index ($\alpha$) and 
normalisation (actually $\log \beta$) even for relatively few periodogram points (i.e. short time series).

The log periodogram regression method has the advantage of being extremely simple 
computationally, so that estimates of the power law parameters (and their 
uncertainties) can be found with minimal effort. However, the method does not
easily generalise to other model forms and does not give the same results as direct 
maximum likelihood analysis\footnote{\citet{Andersson02} provided a modification
of the \citet{Geweke83} fitting method based on the fact that
the logarithm of the periodogram ordinates follow a Gumbel distribution. He gives
the log likelihood function for the logarithm of the periodogram fitted with a
linear function. Maximising this function should give the maximum likelihood
estimates of the power law parameters.} even in the special case of a power law model.

As discussed in \citet{Anderson90}, and
also Appendix A of \citetalias{Vaughan05}, maximum likelihood estimates (MLEs) of the
parameters of a model $S(\bt)$ may be found by maximizing the joint likelihood function
\begin{equation}
\label{eqn:like}
p( \mathbf{I} | \bt, H ) = \prod_{j=1}^{N/2} p(I_j | S_j)
\end{equation}
(cf. eqn \ref{eqn:pdist}),
or equivalently minimising the following function
\begin{equation}
\label{eqn:mlogl}
  D( \mathbf{I}, \bt, H)  =  -2 \log p( \mathbf{I} | \bt, H) 
              =  2 \sum_{j=1}^{N/2} \left\{ \frac{I_j}{S_j} + \log S_j \right\},
\end{equation}
which is twice the minus log likelihood\footnote{
The periodogram is $\chi^2_2$ distributed (equation~\ref{eqn:pdist}) for Fourier frequencies $j=1, 2, \ldots, N/2-1$.
At the Nyquist frequency ($j=N/2$) it has a $\chi^2_1$ distribution.
One could choose to ignore the
Nyquist frequency (sum over $j=1,\ldots,N/2-1$ only), or modify the likelihood function to account for this.
But in the limit of large $N$ the effect on the overall likelihood should be negligible, and so we ignore it here 
and sum over all non-zero Fourier frequencies.}.
This is sometimes known as the Whittle likelihood method, after \citet{Whittle53} and \citet{Whittle57}, and 
has been discussed in detail elsewhere \citep[e.g.][]{Hannan73, Pawitan94, Fan04, Contreras06}.
Here we use the notation $D( \mathbf{I}, \bt)$ for consistency with \citet[][section 6.7]{Gelman04} 
where it is used as the \emph{deviance}, a generalisation of the common
weighted square deviation (or chi square) statistic. Finding the MLEs of the parameters
is the same as finding\footnote{For a function $f(x)$, $\operatorname{arg\,min} f(x)$ gives the the set of points 
of $x$ for which $f(x)$ attains its minimum value.}
\begin{eqnarray}
\label{eqn:mle}
 \hat{\bt}_{\rm MLE} & = & \underset{\bt}{\operatorname{arg\,min}}  ~ D( \mathbf{I}^{\rm obs}, \bt, H) \nonumber \\
                     & = & \underset{\bt}{\operatorname{arg\,max}}  ~ p( \mathbf{I} = \mathbf{I}^{\rm obs}| \bt, H) 
\end{eqnarray}

%__________________________________________________________________

\section{Bayesian periodogram analysis through MCMC}
\label{sect:ba}

We have now laid the groundwork for a fully Bayesian periodogram analysis.
Equation~\ref{eqn:like} gives the likelihood function for the data
given the model $S(\bt)$, or equivalently, equation~\ref{eqn:mlogl} gives the minus log 
likelihood function, which is often easier to work with. 
Once we assign a prior distribution on the model
parameters we can obtain their joint posterior distribution using Bayes theorem (eqn~\ref{eqn:bayes_eqn})
\begin{equation}
 p(\bt | \mathbf{I}, H) \propto p(\mathbf{I} | \bt, H) p(\bt | H) = q(\bt | \mathbf{I}, H).
\end{equation}
where $q(\bt | \mathbf{I}, H)$ is the unnormalised (joint posterior) density function
(the normalisation does not depend on $\bt$).
This can be summarised by the posterior mean (or mode) and credible intervals (equations 
\ref{eqn:post-mean} and \ref{eqn:ci}).
We may also assess the overall fit using a posterior predictive $p$-value for some
useful test quantity.

We may now write an expression for the joint posterior density  (up to a normalisation 
term), or its negative logarithm (up to an additive constant)
\begin{equation}
\label{eqn:mlogpost}
  - \log q(\bt | \mathbf{I}, H) = D( \mathbf{I}, \bt, H)/2 - \log p(\bt|H)
\end{equation}
The posterior mode may then be found by minimising this function (e.g. using a good numerical non-linear minimisation algorithm, 
or Monte Carlo methods in the case of complex, multi-parameter models). 

In the limit of large sample size ($N \rightarrow \infty$) the posterior density will tend to a 
multivariate Normal under quite general conditions \citep[see chapter 4 of][]{Gelman04}. 
For finite $N$ we may make a first approximation to the posterior using a multivariate Normal distribution centred on the mode
and with a covariance matrix $\Sigma$ equal to the curvature of the log posterior at the mode 
(see \citealt{Gelman04}, section 12.2 and \citealt{Albert07}, section 5.5).
(Approximating the posterior as a Normal in this way is often called the \emph{Laplace approximation}.)
This can be used as the basis of a \emph{proposal distribution} in a Markov chain Monte Carlo (MCMC) algorithm
that can efficiently generate draws from the posterior distribution $q(\bt | \mathbf{I} , H)$, given some data
$\mathbf{I} = \mathbf{I}^{\rm obs}$. 
The MCMC was generated by a random-walk Metropolis-Hastings algorithm using a multivariate Normal (with the covariance matrix
as above, but centred on the most recent iteration) as the proposal distribution. 
More details on posterior simulation using MCMC is given in Appendix \ref{sect:mc}.
For each set of simulated parameters we may generate the corresponding spectral model $S(\bt)$ and use this to generate
a periodogram $\mathbf{I}^{\rm rep}$ from the posterior predictive distribution (which in turn may be used to generate 
a time series if needed, see Appendix \ref{sect:sim-ts}).

%______________________________________________________________

\section{Posterior predictive periodogram checks}
\label{sect:ppper}

With the data simulated from the posterior predictive distribution, $\mathbf{I}^{\rm rep}$, we may calculate the distribution
of any test statistic. Of course, we wish to use statistics that are sensitive to the kinds of model deficiency we are interested 
in detecting, such as breaks/bends in the smooth continuum, and narrow peaks due to 
QPOs. Given the arguments of sections \ref{sect:pc} a sensible choice of statistic for investigating QPOs
is $T_{\rm R} = \max_j \hat{R}_j$ (see equation \ref{eqn:Rstat}). Notice that there is no need to perform a multiple-trial 
(Bonferroni) correction to account for the fact that many frequencies are tested before the strongest candidate is selected, as long as this
exact procedure is also applied to the simulated data as the real data.

Another useful statistic is based on the traditional $\chi^2$ statistic, i.e. the sum of the squared standard errors
\begin{equation}
\label{eqn:sse}
 \chi^2(\mathbf{I},\bt) =  \sum_{j=1}^{N/2} \frac{ ( I_j - \E[I_j|\bt] )^2 }{\V[I_j|\bt]} 
             =  \sum_{j=1}^{N/2} \left( \frac{ I_j - S_j(\bt) }{ S_j(\bt) } \right)^2
\end{equation}
where $\E[\cdot]$ and $\V[\cdot]$ indicate expectation and variance, respectively.
We use $T_{SSE} = \chi^2(\mathbf{I}, \hat{\bt})$ where $\hat{\bt}$ is the mode of the posterior distribution.
This is an ``omnibus'' test of the overall data-model match (``goodness-of-fit'') and will be more sensitive to 
inadequacies in the continuum modelling since all data points are included (not just the largest outlier as in $T_{\rm R}$).
This is the same as the merit function used by \citet[][eqn. 16]{Anderson90}, which we call $T_{\rm SSE}$ (for Summed Square Error).

The above two statistics are useful for assessing different aspects of model fitness. By contrast
the Likelihood Ratio Test (LRT) statistic \citep{Eadie71, Cowan98, Protassov02} is a standard tool for comparing 
nested models. As such it may be used to select a continuum model prior to investigating the residuals for possible QPOs. 
The LRT statistic is equal to twice the logarithm of the
ratio of the likelihood maxima for the two models, equivalent to the difference between the 
deviance (which is twice the minimum log likelihood) of the two models
\begin{eqnarray}
\label{eqn:lrt}
  T_{\rm LRT} & = & -2 \log \frac{p(\mathbf{I}|\hat{\bt}_{\rm MLE}^0,H_0)}{p(\mathbf{I}|\hat{\bt}_{\rm MLE}^1,H_1)} \nonumber \\
              & = & D_{\rm min}(H_0) - D_{\rm min}(H_1).
\end{eqnarray}
Asymptotic theory shows that, given certain regularity conditions are met, this statistic should be distributed as a 
chi square variable, $T_{\rm LRT} \sim \chi_{\nu}^2$, 
where the number of degrees of freedom $\nu$ is the difference between the number of free parameters in $H_1$ and $H_0$.
When the regularity conditions are not met \citep[see][]{Freeman99, Protassov02, Park08} we do not 
expect the distribution to be that of the asymptotic theory. 
Nevertheless, the LRT is a powerful statistic for comparing models and can be calibrated by posterior predictive simulation, as shown  
by \citet{Protassov02} and \citet{Rubin94}.

%______________________________________________________________

\section{Application to AGN data}
\label{sect:data}

In this section we apply the method detailed above to two example datasets, both long observations of nearby, variable Seyfert 1 galaxies, obtained from the \xmm\ Science Archive\footnote{See {\tt http://xmm.esac.esa.int/}.}.

\subsection{The power spectrum model}

We shall restrict ourselves to two simple models for the high frequency power spectrum of the Seyferts. 
The first ($H_0$) is a power law plus a constant (to account for the Poisson noise
in the detection process)
\begin{equation}
\label{eqn:m0}
  S(f) = \beta f^{-\alpha} + \gamma
\end{equation}
with three parameters $\bt = \{\alpha, \beta, \gamma \}$, where $\beta$ (the power law normalisation) and $\gamma$ (the additive constant) are constrained to be
non-negative. The second model ($H_1$) is a bending power law as 
advocated by \citet{Mchardy04}
\begin{equation}
\label{eqn:m1}
  S(f) = \beta f^{-1} \left( 1 + \left\{ \frac{f}{\delta} \right\}^{\alpha-1} \right)^{-1} + \gamma
\end{equation}
with four parameters $\bt = \{ \alpha, \beta, \gamma, \delta \}$. For this model $\beta$, $\gamma$ and $\delta$ (the bending
frequency) are all non-negative. The parameter $\alpha$ gives the slope at high frequencies ($f \gg \delta$) in model $H_1$, 
and the low frequency slope is assumed to be $-1$. (This assumption simplifies the model fitting process, and seems reasonable given the
results of \citealt{Uttley02, Markowitz03, Mchardy04} and \citealt{Mchardy06}, but could be relaxed if the model checking process indicated a significant model misfit.)
In the limit of $\delta \rightarrow 0$ the form of $H_1$ tends to that of the simple power law $H_0$.

Following the advice given in \citet{Gelman04} we apply a logarithmic transformation to the non-negative parameters. 
The motivation for this is that the posterior 
should be more symmetric (closer to Normal), and so easier to summarise and handle in computations, 
if expressed in terms of the transformed parameters.
We assign a uniform (uninformative) prior density\footnote{Although strictly speaking these prior densities 
are improper, meaning they do not integrate to unity, we may easily define the prior density to 
be positive only within some large but reasonable range of parameter values, and zero elsewhere, and thereby arrive
at a proper prior density. In the limit of large $N$ the likelihood will dominate over the 
uninformative prior and hence the exact form of the prior density will become irrelevant to the
posterior inferences.} to the transformed parameters, e.g. $p(\alpha, \log \beta, \log \gamma) = const$ for model $H_0$. 
This corresponds to a uniform prior density on the slope $\alpha$ and a 
Jeffreys prior on the parameters restricted to be non-negative (e.g. $p(\beta) = 1/\beta$), which is the conventional prior for
a scale factor \citep{Lee89, Sivia96, Gelman04, Gregory05, Albert07}.

  \begin{figure}
   \centering
   \includegraphics[width=6.5cm,angle=270]{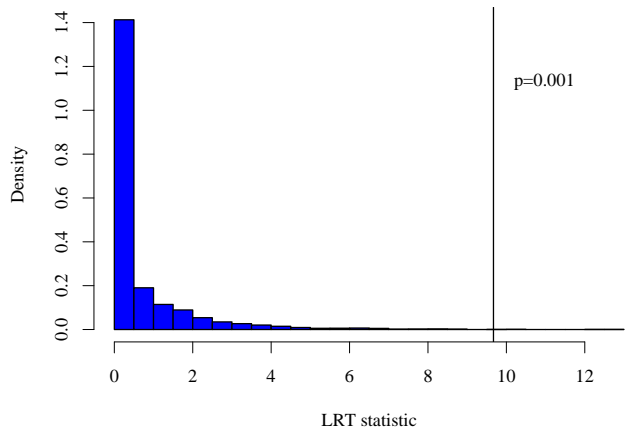}
      \caption{Posterior predictive distribution of the LRT statistic under $H_0$ for the RE J$1034+396$ data 
               (computed using $5,000$ datasets simulated under $H_0$). 
               The observed value $T_{\rm LRT}^{\rm obs}=9.67$ is shown with the vertical line, alongside
               the corresponding $p$-value.
               The distribution is not $\chi^2_1$ as would be predicted by the standard theory 
               but instead resembles a mixture of distributions with half the probability in 
               a $\chi^2_1$ distribution and half concentrated around zero. This might be expected given the
               arguments of \citet[][section 5.4]{Titterington85} and \citet[][Appendix B]{Protassov02}.
              }
         \label{fig:lrt}
   \end{figure}

\begin{table}
\caption{Posterior summaries of parameters for model $H_1$ for the RE J$1034+396$ data. The four parameters are as follows:
              $\alpha =$ power law index, 
              $\beta =$ normalisation (in power density units at $1$ Hz, i.e. $[$rms$/$mean$]^2$ Hz$^{-1}$), 
              $\gamma$ (Poisson noise level in power density units, $[$rms$/$mean$]^2$ Hz$^{-1}$), 
              $\delta$ (bend frequency in Hz). The columns give the parameter name, the posterior mean and the lower 
              and upper bounds of the $90$ per cent credible intervals.}
\label{tab:h1}      % is used to refer this table in the text
\centering                          % used for centering table
\begin{tabular}{l l l l}        % centered columns (4 columns)
\hline\hline                
Parameter      & mean                  & $5$\%                 & $95$\%  \\    
\hline                        
 $\alpha$      & $3.4$                 & $2.2$                 & $5.2$    \\     
 $\beta$       & $2.3 \times 10^{-3}$  & $1.4 \times 10^{-3}$  & $3.7 \times 10^{-3}$ \\
 $\gamma$      & $0.40$                & $0.34$                & $0.45$    \\
 $\delta$      & $4.3 \times 10^{-4}$  & $2.0 \times 10^{-4}$  & $6.5 \times 10^{-4}$  \\
\hline                                  
\end{tabular}
\end{table}

\subsection{Application to \xmm\ data of RE J1034+396}
\label{sect:rej1034}

The first test case we discuss is the interesting \xmm\ observation of the ultrasoft Seyfert 1 galaxy RE J$1034+396$.
\citet{Gierlinski08} analysed these data and reported the detection of a significant QPO which, if confirmed in repeat
observations and by independent analyses, would be the first robust detection of its kind.
For the present analysis a $0.2-10$ keV time series was extracted from the archival data using standard methods \citep[e.g.][]{Vaughan03b}
and binned to $100$ s, to match that used by \citet{Gierlinski08}.

The two candidate continuum models discussed above, $H_0$ and $H_1$ were compared to the data, which gave
$D_{\rm min}^{\rm obs}(H_0) = 504.89$ and $D_{\rm min}^{\rm obs}(H_1) = 495.22$, therefore $T_{\rm LRT}^{\rm obs} = 9.67$.
The MCMC was used to draw from the posterior of model $H_0$, and these draws were used to generate posterior predictive 
periodogram data, which were also fitted with the two models and the results used to map out the posterior predictive
distribution of $T_{\rm LRT}$, which  is shown in Fig. \ref{fig:lrt}. The corresponding tail
area probability for the observed value is $p = 0.001$, small enough that 
the observed reduction in $D_{\rm min}$ between $H_0$ and $H_1$
is larger than might be expected by chance if $H_0$ were true. We therefore favour $H_1$ and use this as the
continuum model. In the absence of complicating factors (see below) this amounts to a significant detection of a power
spectral break.

\begin{figure*}
   \centering
   \includegraphics[width=9cm,angle=270]{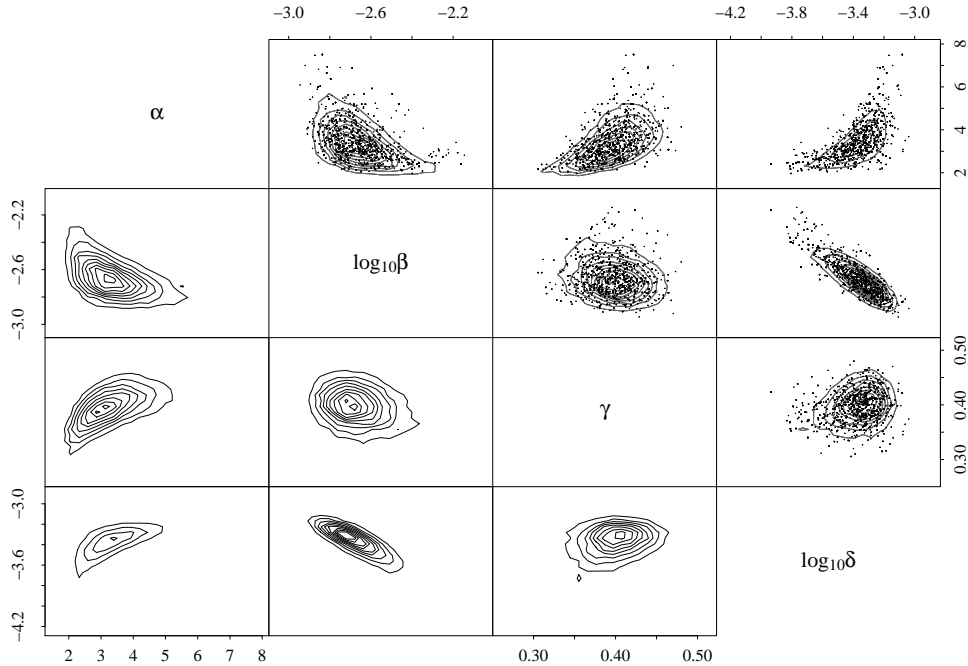}
      \caption{Pairwise marginal posterior distributions for the parameters of $H_1$: 
              $\alpha =$ power law index, 
              $\beta =$ normalisation (in power density units at $1$ Hz, i.e. $[$rms$/$mean$]^2$ Hz$^{-1}$), 
              $\gamma$ (Poisson noise level in power density units, $[$rms$/$mean$]^2$ Hz$^{-1}$), 
              $\delta$ (bend frequency in Hz). The parameters $\beta$ and $\delta$ are shown on a logarithmic scale.
               The lower-left panels show the contours evaluated using all $75,000$ posterior simulations, 
               and the upper-left panels show some of the simulated posterior data (for clarity only $1,000$ points are shown).
              }
         \label{fig:post}
   \end{figure*}

\begin{figure}
   \centering
   \includegraphics[width=6.0cm,angle=270]{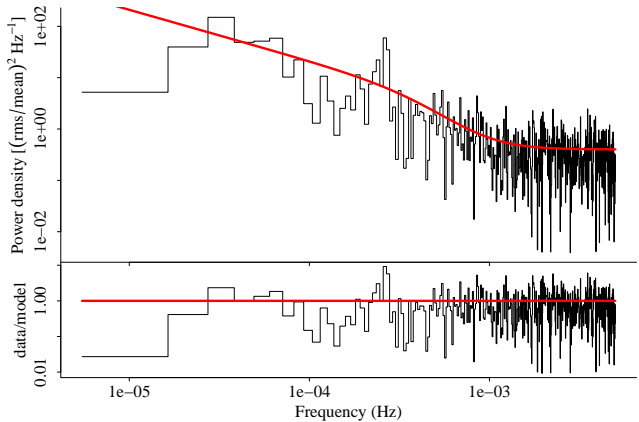}
      \caption{RE J$1034+396$ data and model ($H_1$) computed at the posterior mode. The data are shown as the histogram and 
               the model is shown with the smooth curve. The lower panel shows the data/model residuals
               on a logarithmic scale. See \citet{Gierlinski08} for details of the observation.
              }
         \label{fig:rej-fit}
   \end{figure}

Using $H_1$ as the continuum model we then map out the posterior distribution of the parameters using another MCMC sample. 
Table \ref{tab:h1} presents the posterior means and intervals for the parameters of model $H_1$, and
figure \ref{fig:post} shows the pairwise marginal posterior distributions for the parameters of the model.
Figure \ref{fig:rej-fit} shows the data and model evaluated at the posterior mode.

\begin{figure*}
   \centering
   \hbox{
     \includegraphics[width=6.0cm,angle=270]{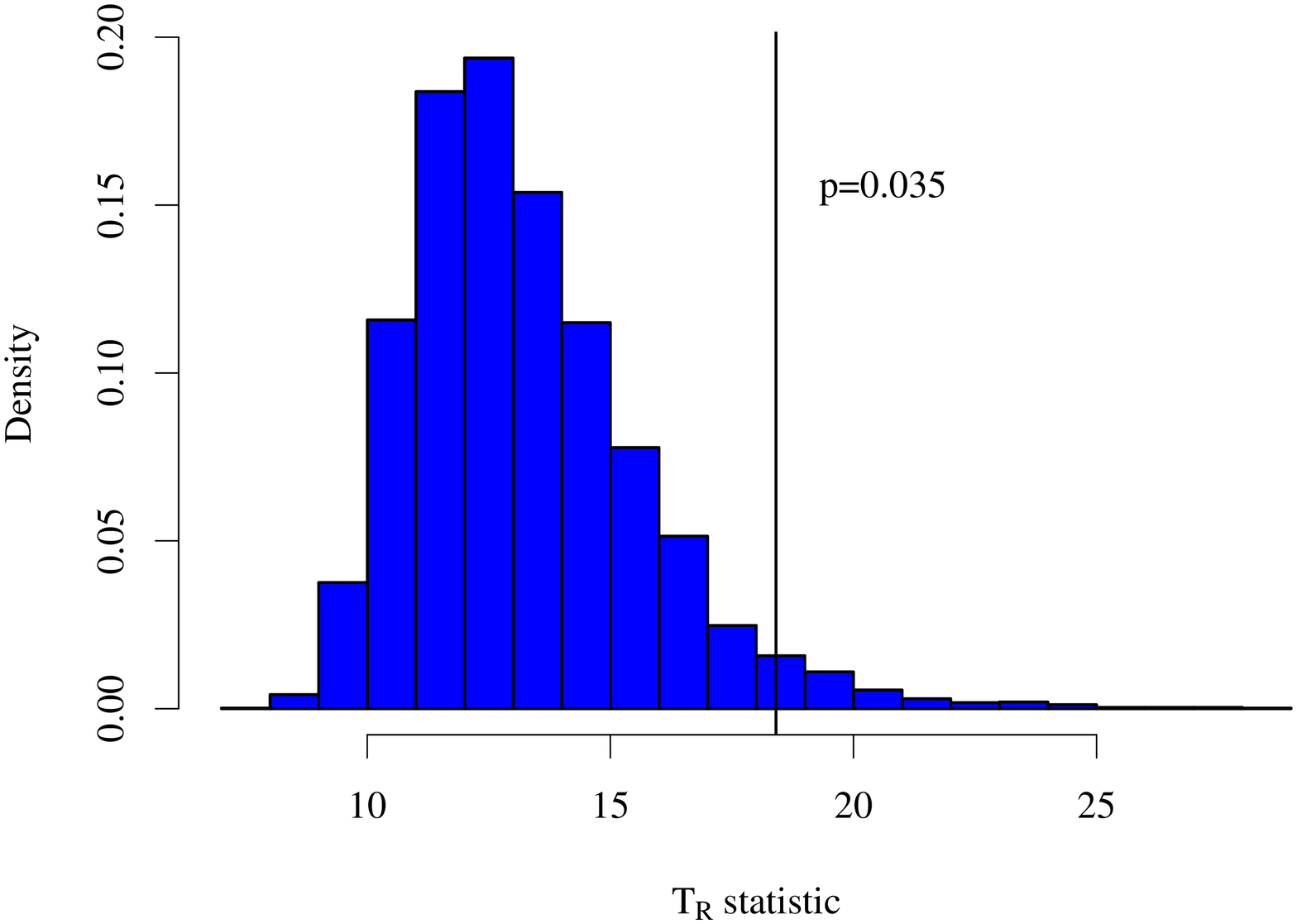}
     \includegraphics[width=6.0cm,angle=270]{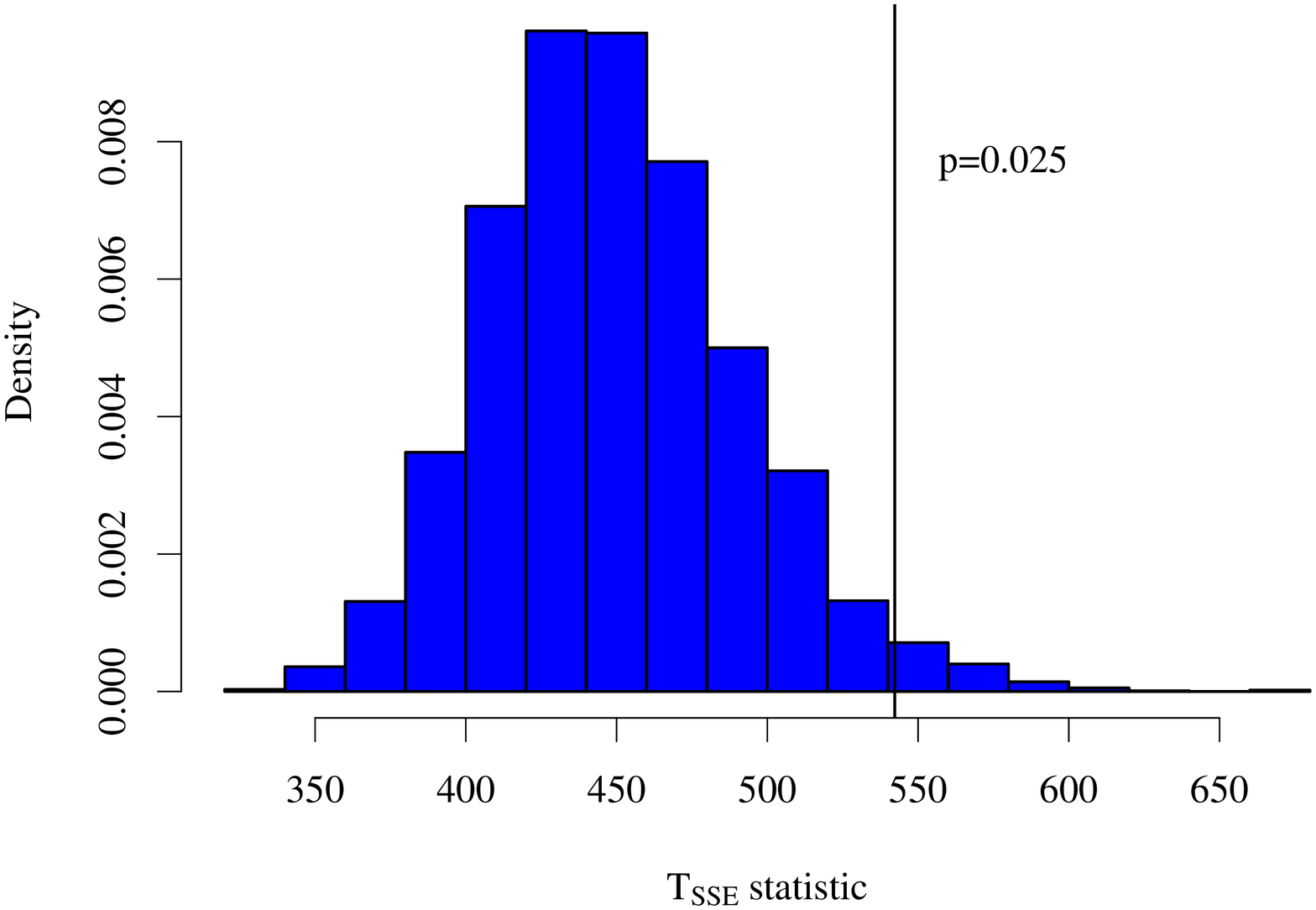}
   }
      \caption{Posterior predictive distributions of the $T_{\rm R}$ and $T_{\rm SSE}$ statistics under $H_1$ for the RE J$1034+396$ data. 
               The observed value of each is shown with a vertical line.
              }
         \label{fig:ppdist}
   \end{figure*}

Clearly there is a large outlier at $\sim 2.5 \times 10^{-3}$ Hz in the residuals after dividing out the model ($H_1$, computed at the posterior mode) which may be due to additional power from a QPO. 
We therefore calculate the posterior predictive distributions of the two test statistics $T_{\rm R}$ and $T_{\rm SSE}$ and compared these to the observed values ($T_{\rm R}^{\rm obs} = 18.41$ and $T_{\rm SSE}^{\rm obs} = 542.3$).
The posterior predictive distributions of these two statistics, derived from $5,000$ simulations, are shown in Fig. \ref{fig:ppdist}. Both these statistics give moderately low $p$-values ($p_{\rm R} = 0.035$ and $p_{\rm SSE} = 0.025$), indicating there is room for improvement in the model and that the largest outlier is indeed rather unusual under $H_1$.
This may indicate the presence of power from a QPO or some other deficiency in the continuum model. Very similar results were obtained after repeating the posterior predictive $p$-value calculations with a variant of $H_1$ in which the low frequency index (at $f \ll \delta$) is fixed at $0$ rather than $-1$, indicating that the $p$-values are not very sensitive to this aspect of the continuum model.

\citet{Gierlinski08} split the time series into two segments and focussed their analysis on the second of these, for which the periodogram residual was 
largest and concentrated in one frequency bin only.
The division of the data into segments is based on a partial analysis of the data -- it is in effect the application of a data-dependent ``stopping rule'' -- and it is extremely difficult to see how such a procedure could be included in the generation of replicated data $\mathbf{I}^{\rm rep}$ used to calibrated the posterior predictive $p$-values. 
We therefore consider $p$-values only for the analysis of the entire time series and do not try to replicate exactly the analysis of \citet{Gierlinski08}.

\subsection{Application to \xmm\ data of Mrk $766$}
\label{sect:mrk766}

A similar analysis was performed on the \xmm\ observation of Mrk 766 discussed previously by
\citet{Vaughan03}, who claimed to have detected a power spectral break using frequentist (classical) statistical tools such as
$\chi^2$ fitting. 
The LRT statistic for the data was $T_{\rm LRT}^{\rm obs} = 18.56$, and the posterior predictive distribution for this statistic
had the same shape as in the case of RE J$1034+396$ (Figure \ref{fig:lrt}).
The $p$-value for the LRT comparison between $H_0$ and $H_1$ was $p < 2 \times 10^{-4}$ (i.e. not one of the $5,000$ 
simulations gave a larger value of $T_{\rm LRT}$). This amounts to a very strong preference for $H_1$ over $H_0$, i.e. a solid detection of a spectral break.

\begin{figure}
   \centering
   \includegraphics[width=6.0cm,angle=270]{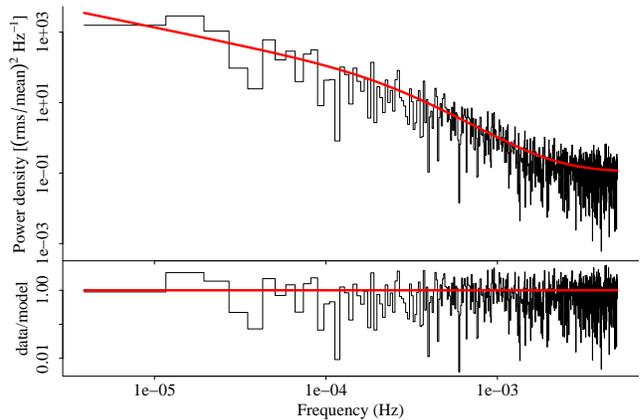}
      \caption{Mrk $766$ data and model ($H_1$) computed at the posterior mode. The panels are the same as in 
      Figure        \ref{fig:rej-fit}.}
         \label{fig:mrk-fit}
   \end{figure}

Table \ref{tab:h2} summarises the posterior inferences for the parameters of $H_1$ and Figure \ref{fig:mrk-fit} shows 
the data, model and residuals. The residuals show no extreme outliers, and indeed the observed values of the test statistics $T_{\rm R}$ and $T_{\rm SSE}$ were not outliers in their posterior predictive distributions ($p_{\rm R}=0.93$ and $p_{\rm SSE}=0.89$). These suggest that $H_1$ provides an adequate description of the data (i.e. without any additional components).

\begin{table}
\caption{Posterior summaries of parameters for model $H_1$ for the Mrk $766$ data. The columns are as in Table~\ref{tab:h1}.}
\label{tab:h2}      % is used to refer this table in the text
\centering                          % used for centering table
\begin{tabular}{l l l l}        % centered columns (4 columns)
\hline\hline                
Parameter      & mean                  & $5$\%                 & $95$\%  \\    
\hline                        
 $\alpha$      & $2.7$                 & $2.4$                 & $3.1$    \\     
 $\beta$       & $1.6 \times 10^{-2}$  & $0.95 \times 10^{-2}$  & $2.7 \times 10^{-2}$ \\
 $\gamma$      & $0.10$                & $0.084$                & $0.12$    \\
 $\delta$      & $2.1 \times 10^{-4}$  & $0.97 \times 10^{-4}$  & $3.4 \times 10^{-4}$  \\
\hline                                  
\end{tabular}
\end{table}

\subsection{Sensitivity to choice of priors}
\label{sect:prior}

It is important to check  the sensitivity of the conclusions to the choice of the prior densities, by studying, for example, the effect of a different or modified choice of prior on the posterior inferences. We have therefore repeated the analysis of the RE J$1034+396$ data using a different choice of priors. In particular, we used independent Normal densities on the four transformed parameters of $H_1$, this is equivalent to a Normal density on the index $\alpha$ and log normal densities on the non-negative valued parameters $\beta$, $\gamma$ and $\delta$. In other words, for each of the transformed parameters $p(\theta_i|H_1) = N(\mu_i, \sigma_i^2)$ where the \emph{hyperparameters} $\mu_i$ and $\sigma_i$ control the mean and width of the prior density functions. After choosing values for the hyperparameters based on knowledge gained from previous studies of nearby, luminous Seyfert galaxies \citep[e.g.][]{Uttley02, Markowitz03, Papadakis04, Mchardy06}, as outlined below, the posterior summaries (parameter means and intervals, pairwise marginal posterior contours, and posterior predictive $p$-values) were essentially unchanged, indicating that the inferences are relatively stable to the choice of prior. 

Previous studies usually gave a high frequency index parameter in the range $\alpha \sim 1-3$, and so we assigned $p(\alpha|H_1) = N(2,4)$, i.e. a prior centred on the typical index of $2$ but with a large dispersion (standard deviation of $2$). 
The normalisation of the $f^{-1}$ part of the power spectrum is thought to be similar between different sources, with $\beta \sim 0.005 - 0.03$ \citep[see][]{Papadakis04}, we assigned $p(\log \beta | H_1) = N(-2,1)$, i.e. a decade dispersion around the mean of $\beta \sim 10^{-2}$. 
The Poisson noise level is dependent on the count rate, which can be predicted very crudely based on previous X-ray observations; we assign a prior $p(\log \gamma | H_1) = N(0,1)$.
The bend/break frequency $\delta$ is thought to correlated with other system parameters such as $M_{\rm BH}$, bolometric luminosity $L_{\rm Bol}$ and optical line width (e.g. $FWHM ~ {\rm H}\beta$). Using the estimated luminosity, and assuming RE J$1034+396$ is radiating close to the Eddington limit \citep{Middleton09} gave a prediction for the bend timescale of $T_{\rm b} \sim 1.6 \times 10^{-3}$ s, and using the optical line width of \citet{Veron01} gave $T_{\rm b} \sim 1.2 \times 10^{-3}$ s, using the relations of \citet{Mchardy06}. Both these (independent) predictions suggest $\delta = 1/T_{\rm b} \sim 10^{-3}$ Hz, and we therefore assigned a prior density $p(\log \delta | H_1) = N(-3,1)$. All of these priors are reasonably non-informative -- they have quite large dispersion around the mean values, to account for the fact that the empirical relations used make these predictions are rather uncertain themselves and also contain intrinsic scatter (i.e. there are significant source to source differences) -- yet they do include salient information about the model obtained from other sources. 

%______________________________________________________________

\section{Discussion}
\label{sect:disco}

We have described, in sections \ref{sect:ml}-\ref{sect:ppper}, a Bayesian analysis of periodogram data that can be
used to estimate the parameters of a power spectral model of a stochastic process, compare two competing continuum
models, and test for the presence of a narrow QPO (or strict periodicity). 

\subsection{Limitations of the method}

The Whittle likelihood function (equation \ref{eqn:like}) is only an approximation to the true sampling distribution of a periodogram.
In the absence of distortions due to the sampling window (more on this below), the ordinates of the periodogram of all stationary, linear 
(and many non-linear) stochastic processes become independently distributed following equation \ref{eqn:pdist} as $N \rightarrow \infty$. 
With finite $N$ (i.e. for real data) this is only approximately true, although with reasonable sample sizes (e.g. $N > 100$) it is a very good approximation.

More serious worries about the distribution of the periodogram, and hence the validity of the Whittle likelihood, come from
distortions due to the sampling effects known as aliasing and leakage \citep[e.g.][]{Uttley02}. It is fairly well established that X-ray light curves from
Seyfert galaxies are stationary once allowance has been made for their red noise character and the linear ``rms-flux'' relation \citep[see][]{Vaughan03b, Uttley05}.
Distortions in the expectation of the periodogram can be modelled by simulating many time series for a given power spectral model, 
resampling these in time as for the original data, and then calculating the average of their periodograms \citep[][and Appendix \ref{sect:sim-ts}]{Uttley02}. 
This does not account for distortions in the distribution of the periodogram ordinates (away from equation \ref{eqn:pdist} predicted by asymptotic theory), 
which is a more challenging problem with (as yet) no accepted solution. However, these affects will be minor or negligible for the 
data analysed in section \ref{sect:data} which are contiguously binned, as the effect of aliasing will be lost in the Poisson noise spectrum which dominates
at high frequencies \citep{vanderklis89, Uttley02}, and the leakage of power from lower to higher frequencies is very low in cases where the power spectrum index is $\alpha \lsim 1$ at the
lowest observed frequencies. The task of fully accounting for sampling distortions in both the expectation and distribution of the periodogram, and hence
having a more general likelihood function, is left for future work.

We should also point out that the usual limitations on the use and interpretation of the periodogram apply. These include the 
(approximate) validity of the Whittle likelihood only when the time series data are evenly sampled. It may be possible to adjust the likelihood function to 
account for the non-independence of ordinates in the modified periodogram usually used with
unevenly sampled time series \citep[e.g.][]{Scargle82}, but here we consider only evenly sampled data.
It is also the case that the periodogram, based on a decomposition of the time series into sinusoidal components, is most sensitive to sinusiodal oscillations, especially 
when they lie close to a Fourier frequency (i.e. the time series spans an integer number of cycles; see \citealt{vanderklis89}).
In situations where the time series is large and spans many cycles of any possible periods (the large $N$ regime), there is no reason to go beyond the standard tools of 
time series processing such as the (time and/or frequency) binned periodogram with approximately normal error bars \citep{vanderklis89}. The current method uses the raw periodogram of a single time series (with the Whittle likelihood) in order to preserve the frequency resolution and bandpass of
the data, which is more important in the low $N$ regime (e.g. when only a few cycles of a suspected period are observed).

The time series data analysed in section \ref{sect:data} were binned up to $100$ s prior to computing the periodogram; this in effect 
ignores frequencies above $5 \times 10^{-3}$ Hz which are sampled by the raw data from the detectors (recorded in counts per CCD frame
at a much higher rate). The choice of bin size does affect the sensitivity to periodic signals of the method described in sections 
\ref{sect:ml}-\ref{sect:ppper}.
Obviously one looses sensitivity to periodic components at frequencies higher than the Nyquist frequency.  
But also as more frequencies are included in the analysis there are more chances to find high $T_{\rm R}$ values from each
simulation, which means the posterior predictive distribution of the test statistic does depend on the choice of binning. 

One could mitigate against this by imposing a priori restrictions on the frequencies of any allowed periods,
for example by altering the test statistic to be $T_{\rm R} = \max_{j < J_0} R_j$ where $J_0$ is some upper limit.
(The lower frequency of the periodogram is restricted by the duration of the time series, which is often dictated by
observational constraints.) But these must be specified independently of the data, otherwise this is in effect another
data-dependent stopping rule (the effect of limiting the frequency range of the search is illustrated below in the 
case of the RE J$1034+396$). 
This sensitivity to choice of binning could be handled more effectively by considering the full frequency range of the periodogram
(i.e. no rebinning of the raw data) and explicitly modelling the periodic component of
the spectrum with an appropriate prior on the frequency range (or an equivalent modelling procedure in the time domain).
But this suffers from the practical drawbacks discussed below.

\subsection{Alternative approaches to model selection}

In many settings the Likelihood Ratio Test (LRT, or the closely related $F$-test) is used to choose between two competing models: the
observed value of the LRT statistic is compared to its theoretical sampling (or reference) distribution, and this is usually summarised with a tail area probability, or $p$-value. 
As discussed above this 
procedure is not valid unless certain specific conditions are satisfied by the data and models. In the case of comparing a single
power law ($H_0$ of section \ref{sect:data}) to a bending power law ($H_1$) the simpler model is reproduced by setting the
extra parameter $\delta \rightarrow 0$ in the more complex model, which violates one of the conditions required by the LRT 
(namely that null values of the extra parameters should not lie at the boundaries of the parameter space). 
In order to use the LRT we must find the distribution of the statistic appropriate for the given data and models, which 
can be done using posterior predictive simulations. This method has the benefit of naturally accounting for nuisance parameters
by giving the expectation of the classical $p$-value over the posterior distribution of the (unknown) nuisance parameters.

One could in principle use the posterior predictive checks to compare a continuum only model (e.g. $H_0$ or $H_1$) to a continuum plus line (QPO) model ($H_2$)
and thereby test for the presence of an additional QPO. \citet{Protassov02} and \citet{Park08} tackled just this problem 
in the context of X-ray energy spectra with few counts. However, we deliberately do not define and use a model with an additional line for the
following reasons. Firstly, this would require a specific line model and a prior density on the line parameters, and it is hard to imagine
these being generally accepted. Unless the line signal is very strong the resulting posterior inferences may be more sensitive to the (difficult) choice
of priors than we would generally wish. Secondly, as shown by \citet{Park08}, there are considerable computational difficulties when using models
with additional, narrow features and data with high variance (as periodograms invariably do), due to the highly multi-modal structure of the likelihood function. 
Our pragmatic alternative is to leave the 
continuum plus line model unspecified, but instead choose a test statistic that is particularly sensitive to narrow excesses in power such as might
be produced under such a model \citep[see][and associated discussions, for more on the choice of test statistic in identifying model deficiency]{Gelman96}.
This has the advantages of not requiring us to specify priors on the line parameters and simplifying the computations, but means the
test is only sensitive to specific types of additional features that have a large effect on the chosen test statistic.
(It is also worth pointing out that the periodogram ordinates are randomly distributed about the spectrum of the stocastic process $S_R(f)$. If
a deterministic process is also present, e.g. producing a strictly periodic component to the signal, this will not in general follow the same
$\chi^2$ distribution and the Whittle likelihood function would need to be modified in order to explicitly model such processes in the spectral domain.)

One of the most popular Bayesian methods for choosing between competing models is the Bayes factor \citep{Kass95, Carlin00, Gelman04, Lee89}. 
These provide a direct comparison of the weight of evidence in favour of one model compared to its competitor, in terms of the ratios of the
marginal likelihoods for the two models (equation \ref{eqn:oddsratio}).
This may be more philosophically attractive than the posterior predictive model checking approach but in practice suffers from the same problems outlined
above, namely the computational challenge of handling a multi-modal likelihood, and the sensitivity to priors on the line parameters, which 
may be even greater for Bayes factors than other methods \citep[see arguments in][]{Protassov02, Gelman04}.

\subsection{Comparison with \citetalias{Vaughan05}}

\citetalias{Vaughan05} tackled the same problem -- the assessment of red noise spectra and detection of additional periodic 
components from short time series -- using frequentist methods. The method developed in the present paper is
superior in a number of ways. The new method is more general in the sense that the model for the continuum power spectrum (i.e. the ``null hypothesis'' model that contains 
no periodicities) may in principle take any parametric form but was previously restricted to a power law. 
It also provides a natural framework for assessing the validity of the continuum model, which should be a crucial step in assessing the
evidence for additional spectral features (see below). Also, by using the
Whittle likelihood rather than the \citet{Geweke83} fit function,  the new method actually gives smaller mean square errors on the
model parameters \citep[see][]{Andersson02}.

\subsection{Comparison with other time series methods}

Previous work on Bayesian methods for period detection \citep[e.g.][]{Bretthorst88, Gregory92, Gregory99} has focussed on cases
where the stochastic process is assumed to be white (uncorrelated) noise on which a strictly periodic signal is superposed. They do not 
explicitly tackle the more general situation of a non-white continuum spectrum that is crucial to analysing data from
compact accreting X-ray sources.

The only non-Bayesian (i.e. frequentist) methods we are aware of for assessing evidence for periodicities in 
data with a non-white spectrum involve applying some kind of smoothing to the raw periodogram data. This gives a non-parametric estimate of the 
underlying spectrum, with some associated uncertainty on the estimate, which can then be compared to the unsmoothed periodogram data and used to search for outlying periodogram 
points.
The Multi-Taper Method (MTM) of \citet{Thompson82} \citep[also][]{Thompson90} achieves the smoothing by averaging the multiple periodograms, each computed 
using one member of a set of orthogonal data tapers. See \citet[][chapter 7]{Percival93} for a good discussion of this method.
The data tapers are designed to reduce spectral leakage and so reduced bias in the resulting spectrum estimate.
The method proposed by \citet{Israel96} involves a more straightforward running mean of the peridogram data. 
Both of these are non-parametric methods, meaning that they do not involve a specific parametric model for the underlying spectrum.
This lack of model dependence might appear to be an advantage, but in fact may be a disadvantage in cases where we do have good
reasons for invoking a particular type of parametric model (e.g. the bending power laws seen in the Seyfert galaxy data).
The continuum model's few parameters may be well constrained by the data, where the non-parametric (smoothed) estimate at each frequency is not.
The non-parametric methods also leave a somewhat arbitrary choice of how to perform the smoothing, i.e. the type and number of data tapers in the MTM, 
or the size/shape of the smoothing kernel in the \citet{Israel96} method.
Also, it is less obvious how to combine the sampling distribution of the periodogram ordinate (line component) and the spectrum estimate (continuum), 
and how to account for the number of ``independent'' frequencies searched. These are all automatically included in the posterior predictive $p$-value method
as outlined above.

In the present paper we have  deliberately concentrated on the periodogram since this is the 
standard tool for time series analysis in astronomy. But the periodogram is by no means the best or only tool for the 
characterisation of stochastic processes or the identification of periodicities.
Methods that explicitly model the original time series data in the time domain \citep[see e.g.][]{Priestley81, Chatfield03} 
may yet prove to be valuable additions to the astronomers toolkit. 
Indeed the raw form of the \xmm\ data used in the AGN examples is counts per CCD frame, for the source (and possibly background
region if this is a non-negligible contribution). The most direct data analysis would therefore model this process explicitly 
as a Poisson process with a rate parameter that varies with time (i.e. the ``true'' X-ray flux) that is itself a realisation of
some stochastic process with specific properties (e.g. power spectrum or, equivalently, autocorrelation function, and stationary distribution). 

\subsection{The importance of model assessment}

The posterior predictive  approach provides an attractive scheme for model checking. In particular, it allows us to 
select a  continuum model that is consistent with the observed data\footnote{Strictly, we compare the 
observed data to simulations drawn from the posterior predictive distribution under the 
chosen model $H$ using test statistics. If the observed data do not stand out from the simulations, 
by having extreme values of the statistics when compared to the simulations, 
we may assume that the data are consistent with the model (as far as the particular test statistics are concerned).} 
before testing for the presence of additional features. This is crucial since any
simple test statistic, whether used in a frequentist significance test or a posterior predictive test, will be sensitive to 
certain kinds of deficiencies in the model without itself providing any additional information about the specific nature of any deficiency detected
(a $p$-values is after all just a single number summary).
A low $p$-value (i.e. a ``significant'' result) may be due to the presence of interesting additional features or just an overall poor match
between the data and the continuum model \citep[for more on this in the context of QPO detection see][]{Vaughan06}.
The use of more than one test statistic, properly calibrated using the posterior predictive simulations, as well as other model
diagnostics (such as data/model residual plots) are useful in identifying the cause of the data/model mismatch.

\subsection{Analysis of two Seyfert galaxies}

  \begin{figure}
   \centering
   \includegraphics[width=6.0cm,angle=270]{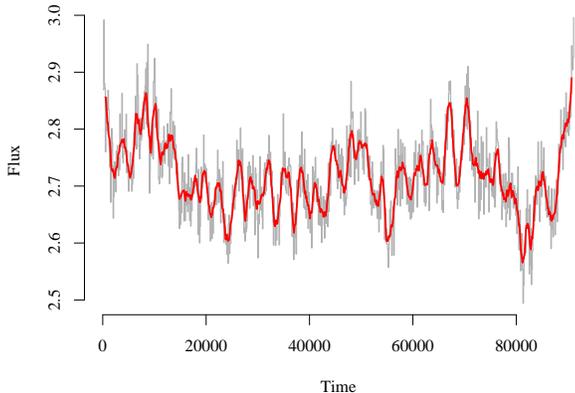}
      \caption{Simulated time series generated from the posterior predictive distribution of the RE J$1034+396$ periodogram data. 
      The (grey) histogram shows the simulated data in $100$ s bins and the smooth (red) curve shows the $6$ bin moving average 
      of these data. Compare with Figure 1 of \citet{Gierlinski08}. The power spectrum used to generate these data is a smoothly
      bending power law (plus white ``measurement'' noise) with no periodic or quasi-periodic components, and they the 
      time series appears to show oscillatory structure.}
         \label{fig:sim}
   \end{figure}

Section \ref{sect:data} presents an analysis of \xmm\ data for the Seyfert galaxies RE J$1034+396$ and Mrk $766$.
The former has produced the best evidence to date for a QPO in a Seyfert galaxy \citep{Gierlinski08}, while
the latter showed no indication of QPO behaviour \citep{Vaughan03, Vaughan05b}. 
\citet{Gierlinski08} used
the method presented in \citetalias{Vaughan05} to show that the observed peak in the periodogram was highly
unlikely under the assumption than the underlying power spectrum continuum is a power law, but the present analysis
gave somewhat less impressive evidence to suggest a QPO.

The posterior predictive $p \approx 0.03$ comes from the fact that $\sim 150$ out of the $5,000$ posterior predictive simulations of
the RE J$1034+396$ periodogram data showed $T_{\rm R} \ge T_{\rm R}^{\rm obs}$ (and approximately the 
same figure was obtained using $T_{\rm SSE}$).
This might at first seem doubtful given how periodic the observed time series appears (see Figure 1 of \citealt{Gierlinski08}).
But to demonstrate that such apparently periodic time series may indeed be generated from non-periodic processes we simulated time
series from the posterior predictive periodogram data (for model $H_1$) that showed $T_{\rm R} \ge T_{\rm R}^{\rm obs}$. 
(The time series simulation method is given in Appendix \ref{sect:sim-ts}.) 
One example of these time series, chosen at random from the subset that had the largest residual $R_j$ occurring at a frequency of the same order as that seen in RE J$1034+396$ (in this case $\approx 1.3 \times 10^{-4}$ Hz), is shown in Figure \ref{fig:sim}.

There are several reasons for the very different $p$-values between the analyses.
One of these factors is that we based our calculation on a more general form of the continuum model.
In the absence of a QPO (spectral line component) the power spectrum continuum is well
modelled using a power law with a steep slope ($\alpha \sim 3$) that smoothly changes to a flatter slope (assumed index of $-1$) 
below a frequency $\delta \sim 4 \times 10^{-4}$ Hz, than a single power law. The bend frequency is close to that of the candidate QPO, which does have 
a large effect on the ``significance'' of the QPO as summarised in the $p$-value \citep[see][for previous examples of this effect]{Vaughan05b}. 
Indeed, the posterior predictive $p$-value was $2 \times 10^{-3}$ when recalculated assuming a simple power law continuum ($H_0$).
A second factor is that \citet{Gierlinski08} gave special consideration to a particular subset of the times series
chosen because of its apparently coherent oscillations, which in effect enhanced the apparent significance of the claimed periodicity,
while the entire time series is treated uniformly in the present analysis (for reasons discussed in section \ref{sect:data}).
A third factor is that we made no restriction on the allowed frequency of a period component, and so 
openly searched $457$ frequencies, where \citet{Gierlinski08} concentrated
on the $\approx 60$ frequencies in their periodogram below $10^{-3}$ Hz. This will result in a factor $\sim 8$ change in the
$p$-value (since the probability of finding a $T_{\rm R}$ value in a simulation that is larger than the observed in the real data scales
approximately linearly with the number of frequencies examined). If we take $T_{\rm R}$ to be the largest residual
at frequencies below $10^{-3}$ Hz (but including all the data in the rest of the modelling process), we find $21/5000$ of the RE J$1034+396$ simulations 
showed $T_{\rm R} \ge T_{\rm R}^{\rm obs}$ under these restricted conditions, corresponding to $p=0.004$, which is smaller by about the
expected factor.
A relatively minor difference is the more complete treatment of 
parameter uncertainties using the posterior distribution \citepalias[which is treated in an approximation fashion in the method of][]{Vaughan05}.
One is therefore left with a choice between two models that could plausibly explain the data, a power law spectrum with a strong QPO or a 
bending power law spectrum (with weaker evidence for a QPO). The most powerful and least ambiguous confirmation of the reality of the QPO
feature would come from a independent observation capable of both constraining the continuum more precisely and allowing a sensitive search for the
candidate QPO.

The results of the present analysis of the  Mrk $766$ data agree reasonably well with those previously reported by \citet{Vaughan03b}
which were obtained using standard frequentist methods (e.g. binning the data and estimating parameters by minimising the $\chi^2$ statistic). 
The high frequency slopes are essentially the same, but the frequency of the bend differs by a factor of $\sim 2.5$. This is most likely
due to the slightly different models used, i.e. bending vs. sharply broken power laws. (Repeating the frequentist analysis of \citet{Vaughan03b} using 
the bending power law model gave a lower characteristic frequency, more consistent with that of the present analysis).

\subsection{Other applications of this method}

The techniques discussed in this paper may find application well beyond the specific field for which they were devised (namely,
analysis of X-ray light curves from Seyfert galaxies), since the problems of estimating a noisy continuum spectrum and 
assessing the evidence for additional narrow features over and above that continuum are common to many fields.  
Other examples from X-ray astronomy include analysis of long timescale light curves from Galactic X-ray binaries and Ultra-Luminous X-ray sources (ULXs) 
in order to characterise the low frequency power spectrum and search for periodicities (e.g. due to orbital modulation).

But the applications are by no means restricted to astronomy. For example, in geology there is considerable interest in detecting
and characterising periodicities in stratigraphic records of environmental change, which may be connected to periodicities in
external forcing such as might be expected from Milankovich cycles \citep[see e.g.][]{Weedon03}. However, there is controversy 
over the statistical and physical significance of the periodicities in these data, which are often dominated by stochastic red noise
variations \citep{Bailey09}.

%______________________________________________________________

\section{Conclusions}
\label{sect:conc}

We have presented Bayesian methods for the modelling of periodogram data that can be used for both 
parameter estimation and model checking, and may be used to test for narrow spectral
features embedded in noisy data. The model assessment is performed using simulations of posterior predictive 
data to calibrate (sensibly chosen) test statistics. This does however leave some arbitrariness in the 
method, particularly in the choice of test statistic\footnote{In situations where two competing models
can be modelled explicitly the LRT provides a natural choice of statistic.} (and in some situations the choice of what 
constitutes a simulation of the data).
Such issues were always present, if usually ignored, in the standard frequentist tests. The posterior predictive
approach has the significant advantage of properly treating nuisance parameters, and provides a clear framework
for checking the different aspects of the reasonableness of a model fit. 
The issue of choosing a test statistic does not arise in more ``purist'' Bayesian methods such as Bayes factors, which 
concentrate on the posterior distributions and marginal likelihoods, but such methods of 
model selection carry their own burden in terms of the computational complexity and the difficulty of selecting 
(and the sensitivity of inferences to) priors on the model parameters. 
The method presented in this paper, making use of the posterior predictive checking, is an improvement over the
currently popular methods that use classical $p$-value; but Bayesian model selection is an area
of active research and it is not unreasonable to expect that new, powerful and practical computational tools 
will be developed or adapted to help bridge the gap between the pragmatic and the purist Bayesian approaches.

The routines used to perform the analysis of the real data presented in section \ref{sect:data} will be
made available as an {\tt R}\footnote{{\tt R} is a powerful, open-source computing environment for data 
analysis and statistics that may be downloaded for free from {\tt http://www.r-project.org/} \citep{r, r2}.}
script from the author upon request.

%______________________________________________________________

\section*{Acknowledgements}
The author wishes to thank David van Dyk and Phil Uttley for valuable discussions during the
final stages of writing this paper, and an anonymous referee for a helpful report.

%______________________________________________________________

\bibliographystyle{mn2e}
\bibliography{aa}

\begin{thebibliography}{}

\bibitem[\protect\citeauthoryear{Albert}{Albert}{2007}]{Albert07}
Albert J.,  2007, Bayesian Computation with R.
Springer, New York

\bibitem[\protect\citeauthoryear{{Anderson}, {Duvall} Jr. \&
  {Jefferies}}{{Anderson} et~al.}{1990}]{Anderson90}
{Anderson} E.~R.,  {Duvall} Jr. T.~L.,    {Jefferies} S.~M.,  1990, \apj, 364,
  699

\bibitem[\protect\citeauthoryear{Andersson}{Andersson}{2002}]{Andersson02}
Andersson J.,  2002, Economics Letters, 77, 137

\bibitem[\protect\citeauthoryear{{Bailey}}{{Bailey}}{2009}]{Bailey09}
{Bailey} R.~J.,  2009, Terra Nova, in press.

\bibitem[\protect\citeauthoryear{{Berger} \& {Berry}}{{Berger} \&
  {Berry}}{1988}]{Berger88}
{Berger} J.~O.,  {Berry} D.~A.,  1988, American Scientist, 76, 159

\bibitem[\protect\citeauthoryear{{Berry}}{{Berry}}{1996}]{Berry96}
{Berry} D.~A.,  1996, Statistics: A Bayesian Perspective.
Duxbury, London

\bibitem[\protect\citeauthoryear{Bloomfield}{Bloomfield}{2000}]{Bloomfield00}
Bloomfield P.,  2000, Fourier Analysis of Time Series: An Introduction, 2nd
  edn.
Wiley, New York

\bibitem[\protect\citeauthoryear{{Bretthorst}}{{Bretthorst}}{1988}]{Bretthorst%
88}
{Bretthorst} G.~L.,  1988, Bayesian Spectrum Analysis and Parameter Estimation.
Lecture Notes in Statistics, Springer-Verlag, Heidelberg

\bibitem[\protect\citeauthoryear{{Carlin} \& {Louis}}{{Carlin} \&
  {Louis}}{2000}]{Carlin00}
{Carlin} B.~P.,  {Louis} T.~A.,  2000, Bayes and Empirical Bayes Methods for
  Data Analysis (2nd ed.).
Chapman \& Hall/CRC, London

\bibitem[\protect\citeauthoryear{Chatfield}{Chatfield}{2003}]{Chatfield03}
Chatfield C.,  2003, The Analysis of Time Series: An Introduction.
{Chapman \& Hall/CRC}, London

\bibitem[\protect\citeauthoryear{Chib \& Greenberg}{Chib \&
  Greenberg}{1995}]{Chib95}
Chib S.,  Greenberg E.,  1995, The American Statistician, 49, 327

\bibitem[\protect\citeauthoryear{{Contreras-Crist\'{a}n},
  {Guti\'{e}rrez-Pe\~{n}a} \& {Walker}}{{Contreras-Crist\'{a}n}
  et~al.}{2006}]{Contreras06}
{Contreras-Crist\'{a}n} E.,  {Guti\'{e}rrez-Pe\~{n}a} E.,    {Walker} S.~G.,
  2006, Communications in Statistics (Simulation and Computation), 35, 857

\bibitem[\protect\citeauthoryear{{Cowan}}{{Cowan}}{1998}]{Cowan98}
{Cowan} G.,  1998, Statistical data analysis.
Clarendon Press, Oxford

\bibitem[\protect\citeauthoryear{Davies \& Harte}{Davies \&
  Harte}{1987}]{Davies87}
Davies R.~B.,  Harte D.~S.,  1987, Biometrica, 74, 95

\bibitem[\protect\citeauthoryear{{Eadie}, {Drijard}, {James}, {Roos} \&
  {Sadoulet}}{{Eadie} et~al.}{1971}]{Eadie71}
{Eadie} W.~T.,  {Drijard} D.,  {James} F.~E.,  {Roos} M.,    {Sadoulet} B.,
  1971, {Statistical methods in experimental physics}.
North-Holland, Amsterdam

\bibitem[\protect\citeauthoryear{Fan \& Zhang}{Fan \& Zhang}{2004}]{Fan04}
Fan J.,  Zhang W.,  2004, Biometrika, 91, 195

\bibitem[\protect\citeauthoryear{{Fisher}}{{Fisher}}{1929}]{Fisher29}
{Fisher} R.~A.,  1929, Proceedings of the Royal Society of London: Series A,
  125, 54

\bibitem[\protect\citeauthoryear{{Freeman}, {Graziani}, {Lamb}, {Loredo},
  {Fenimore}, {Murakami} \& {Yoshida}}{{Freeman} et~al.}{1999}]{Freeman99}
{Freeman} P.~E.,  {Graziani} C.,  {Lamb} D.~Q.,  {Loredo} T.~J.,  {Fenimore}
  E.~E.,  {Murakami} T.,    {Yoshida} A.,  1999, \apj, 524, 753

\bibitem[\protect\citeauthoryear{{Gamerman}}{{Gamerman}}{1997}]{Gamerman97}
{Gamerman} D.,  1997, Markov Chain Monte Carlo: Stochastic Simulation for
  Bayesian Inference.
Chapman and Hall/CRC

\bibitem[\protect\citeauthoryear{{Gelman}, {Carlin}, {Stern} \& B.}{{Gelman}
  et~al.}{2004}]{Gelman04}
{Gelman} A.,  {Carlin} J.~B.,  {Stern} H.~S.,    B. R.~D.,  2004, Bayesian Data
  Analysis (2nd ed).
Chapman \& Hall, London

\bibitem[\protect\citeauthoryear{{Gelman}, {Meng} \& {Stern}}{{Gelman}
  et~al.}{1996}]{Gelman96}
{Gelman} A.,  {Meng} X.-L.,    {Stern} H.~S.,  1996, Statistica Sinica, 6, 733

\bibitem[\protect\citeauthoryear{Gelman \& Rubin}{Gelman \&
  Rubin}{1992}]{Gelman92}
Gelman A.,  Rubin D.~B.,  1992, Statistical Science, 7, 457

\bibitem[\protect\citeauthoryear{{Geweke} \& {Porter-Hudak}}{{Geweke} \&
  {Porter-Hudak}}{1983}]{Geweke83}
{Geweke} J.,  {Porter-Hudak} S.,  1983, Journal of Time Series Analysis, 4, 221

\bibitem[\protect\citeauthoryear{{Gierli{\'n}ski}, {Middleton}, {Ward} \&
  {Done}}{{Gierli{\'n}ski} et~al.}{2008}]{Gierlinski08}
{Gierli{\'n}ski} M.,  {Middleton} M.,  {Ward} M.,    {Done} C.,  2008, \nat,
  455, 369

\bibitem[\protect\citeauthoryear{{Gilks}, {Richardson} \&
  {Spiegelhalter}}{{Gilks} et~al.}{1995}]{Gilks95}
{Gilks} W.~R.,  {Richardson} S.,    {Spiegelhalter} D.,  1995, Markov Chain
  Monte Carlo in Practice.
{Chapman \& Hall/CRC}

\bibitem[\protect\citeauthoryear{{Gregory}}{{Gregory}}{1999}]{Gregory99}
{Gregory} P.~C.,  1999, \apj, 520, 361

\bibitem[\protect\citeauthoryear{{Gregory}}{{Gregory}}{2005}]{Gregory05}
{Gregory} P.~C.,  2005, {Bayesian Logical Data Analysis for the Physical
  Sciences}.
Cambridge University Press, Cambridge, UK

\bibitem[\protect\citeauthoryear{{Gregory} \& {Loredo}}{{Gregory} \&
  {Loredo}}{1992}]{Gregory92}
{Gregory} P.~C.,  {Loredo} T.~J.,  1992, \apj, 398, 146

\bibitem[\protect\citeauthoryear{{Groth}}{{Groth}}{1975}]{Groth75}
{Groth} E.~J.,  1975, \apjs, 29, 285

\bibitem[\protect\citeauthoryear{{Hannan}}{{Hannan}}{1973}]{Hannan73}
{Hannan} E.~J.,  1973, J. Appl. Prob., 10, 130

\bibitem[\protect\citeauthoryear{{Howson} \& {Urbach}}{{Howson} \&
  {Urbach}}{1991}]{Howson91}
{Howson} C.,  {Urbach} P.,  1991, Nature, 350, 371

\bibitem[\protect\citeauthoryear{{Israel} \& {Stella}}{{Israel} \&
  {Stella}}{1996}]{Israel96}
{Israel} G.~L.,  {Stella} L.,  1996, \apj, 468, 369

\bibitem[\protect\citeauthoryear{{Jeffreys} \& {Berger}}{{Jeffreys} \&
  {Berger}}{1992}]{Jeffreys92}
{Jeffreys} W.~H.,  {Berger} J.~O.,  1992, American Scientist, 80, 64

\bibitem[\protect\citeauthoryear{{Jenkins} \& {Watts}}{{Jenkins} \&
  {Watts}}{1969}]{Jenkins69}
{Jenkins} G.~M.,  {Watts} D.~G.,  1969, {Spectral analysis and its
  applications}.
Holden-Day, London

\bibitem[\protect\citeauthoryear{Kass, Carlin, Gelman \& Neal}{Kass
  et~al.}{1998}]{Kass98}
Kass R.~E.,  Carlin B.~P.,  Gelman A.,    Neal R.~M.,  1998, The American
  Statistician, 52, 93

\bibitem[\protect\citeauthoryear{{Kass} \& {Raftery}}{{Kass} \&
  {Raftery}}{1995}]{Kass95}
{Kass} R.~E.,  {Raftery} A.~E.,  1995, J. Am. Stat. Ass., 90, 773

\bibitem[\protect\citeauthoryear{{Leahy}, {Darbro}, {Elsner}, {Weisskopf},
  {Kahn}, {Sutherland} \& {Grindlay}}{{Leahy} et~al.}{1983}]{Leahy83}
{Leahy} D.~A.,  {Darbro} W.,  {Elsner} R.~F.,  {Weisskopf} M.~C.,  {Kahn} S.,
  {Sutherland} P.~G.,    {Grindlay} J.~E.,  1983, \apj, 266, 160

\bibitem[\protect\citeauthoryear{{Lee}}{{Lee}}{2004}]{Lee89}
{Lee} P.~M.,  2004, {Bayesian Statistics: An Introduction (3rd ed)}.
Wiley, New York

\bibitem[\protect\citeauthoryear{{Loredo}}{{Loredo}}{1990}]{Loredo90}
{Loredo} T.~J.,  1990, in Fougere P.,  ed., , Maximum-Entropy and Bayesian
  Methods, Dartmouth..
Kluwer Academic Publishers, Dordrecht, The Netherlands, pp 81--142

\bibitem[\protect\citeauthoryear{{Loredo}}{{Loredo}}{1992}]{Loredo92}
{Loredo} T.~J.,  1992, in Feigelson D.,  Babu G.,  eds, , Statistical
  Challenges in Modern Astronomy, Springer-Verlag..
Springer-Verlag, New York, pp 275--297

\bibitem[\protect\citeauthoryear{{Markowitz}, {Edelson}, {Vaughan}, {Uttley},
  {George}, {Griffiths}, {Kaspi}, {Lawrence}, {McHardy}, {Nandra}, {Pounds},
  {Reeves}, {Schurch} \& {Warwick}}{{Markowitz} et~al.}{2003}]{Markowitz03}
{Markowitz} A.,  {Edelson} R.,  {Vaughan} S.,  {Uttley} P.,  {George} I.~M.,
  {Griffiths} R.~E.,  {Kaspi} S.,  {Lawrence} A.,  {McHardy} I.,  {Nandra} K.,
  {Pounds} K.,  {Reeves} J.,  {Schurch} N.,    {Warwick} R.,  2003, \apj, 593,
  96

\bibitem[\protect\citeauthoryear{{McHardy}, {Koerding}, {Knigge}, {Uttley} \&
  {Fender}}{{McHardy} et~al.}{2006}]{Mchardy06}
{McHardy} I.~M.,  {Koerding} E.,  {Knigge} C.,  {Uttley} P.,    {Fender} R.~P.,
   2006, \nat, 444, 730

\bibitem[\protect\citeauthoryear{{McHardy}, {Papadakis}, {Uttley}, {Page} \&
  {Mason}}{{McHardy} et~al.}{2004}]{Mchardy04}
{McHardy} I.~M.,  {Papadakis} I.~E.,  {Uttley} P.,  {Page} M.~J.,    {Mason}
  K.~O.,  2004, \mnras, 348, 783

\bibitem[\protect\citeauthoryear{Meng}{Meng}{1994}]{Meng94}
Meng X.-L.,  1994, Annals of Statistics, 22, 1142

\bibitem[\protect\citeauthoryear{{Middleton}, {Done}, {Ward}, {Gierli{\'n}ski}
  \& {Schurch}}{{Middleton} et~al.}{2009}]{Middleton09}
{Middleton} M.,  {Done} C.,  {Ward} M.,  {Gierli{\'n}ski} M.,    {Schurch} N.,
  2009, \mnras, 394, 250

\bibitem[\protect\citeauthoryear{{Papadakis}}{{Papadakis}}{2004}]{Papadakis04}
{Papadakis} I.~E.,  2004, \mnras, 348, 207

\bibitem[\protect\citeauthoryear{{Park}, {van Dyk} \& {Siemiginowska}}{{Park}
  et~al.}{2008}]{Park08}
{Park} T.,  {van Dyk} D.~A.,    {Siemiginowska} A.,  2008, \apj, 688, 807

\bibitem[\protect\citeauthoryear{Pawitan \& O'Sullivan}{Pawitan \&
  O'Sullivan}{1994}]{Pawitan94}
Pawitan Y.,  O'Sullivan F.,  1994, Journal of the American Statistical
  Association, 89, 600

\bibitem[\protect\citeauthoryear{Percival \& Walden}{Percival \&
  Walden}{1993}]{Percival93}
Percival D.~B.,  Walden A.~T.,  1993, Spectral analysis for physical
  applications : multitaper and conventional univariate techniques.
Cambridge University Press, Cambridge

\bibitem[\protect\citeauthoryear{{Pilgram} \& {Kaplan}}{{Pilgram} \&
  {Kaplan}}{1998}]{Pilgram98}
{Pilgram} B.,  {Kaplan} D.~T.,  1998, Phys. D, 114, 108

\bibitem[\protect\citeauthoryear{{Press}}{{Press}}{1978}]{Press78}
{Press} W.~H.,  1978, Comments on Astrophysics, 7, 103

\bibitem[\protect\citeauthoryear{{Press}, {Teukolsky}, {Vetterling} \&
  {Flannery}}{{Press} et~al.}{1992}]{Press92}
{Press} W.~H.,  {Teukolsky} S.~A.,  {Vetterling} W.~T.,    {Flannery} B.~P.,
  1992, {Numerical recipes in FORTRAN. The art of scientific computing}.
Cambridge: University Press, |c1992, 2nd ed.

\bibitem[\protect\citeauthoryear{{Priestley}}{{Priestley}}{1981}]{Priestley81}
{Priestley} M.~B.,  1981, Spectral Analysis and Time Series.
Academic Press, London

\bibitem[\protect\citeauthoryear{{Protassov}, {van Dyk}, {Connors}, {Kashyap}
  \& {Siemiginowska}}{{Protassov} et~al.}{2002}]{Protassov02}
{Protassov} R.,  {van Dyk} D.~A.,  {Connors} A.,  {Kashyap} V.~L.,
  {Siemiginowska} A.,  2002, \apj, 571, 545

\bibitem[\protect\citeauthoryear{{R Development Core Team}}{{R Development Core
  Team}}{2009}]{r}
{R Development Core Team} 2009, R: A Language and Environment for Statistical
  Computing.
http://www.R-project.org, Vienna, Austria

\bibitem[\protect\citeauthoryear{Rubin}{Rubin}{1984}]{Rubin84}
Rubin D.~B.,  1984, Annals of Statistics, 12, 1151

\bibitem[\protect\citeauthoryear{{Rubin} \& {Stern}}{{Rubin} \&
  {Stern}}{1994}]{Rubin94}
{Rubin} D.~B.,  {Stern} H.~S.,  1994, in {von Eye} A.,  {Clogg} C.,  eds,
  Latent Variables Analysis: Applications for Developmental Research {Testing
  in Latent Class Models using a Posterior Predictive Check Distribution}.
pp 420--438

\bibitem[\protect\citeauthoryear{{Scargle}}{{Scargle}}{1982}]{Scargle82}
{Scargle} J.~D.,  1982, \apj, 263, 835

\bibitem[\protect\citeauthoryear{{Sivia}}{{Sivia}}{1996}]{Sivia96}
{Sivia} D.~S.,  1996, {Data Analysis: A Bayesian Tutorial}.
Oxford Univ. Press, Oxford

\bibitem[\protect\citeauthoryear{Thompson}{Thompson}{1990}]{Thompson90}
Thompson D.~J.,  1990, Phil. Trans. R. Soc. Lond. A, 332, 539

\bibitem[\protect\citeauthoryear{Thomson}{Thomson}{1982}]{Thompson82}
Thomson D.~J.,  1982, Proceedings of the IEEE, 70, 1055

\bibitem[\protect\citeauthoryear{Tierney}{Tierney}{1994}]{Tierney94}
Tierney L.,  1994, The Annals of Statistics, 22, 1701

\bibitem[\protect\citeauthoryear{{Timmer} \& {K{\"o}nig}}{{Timmer} \&
  {K{\"o}nig}}{1995}]{Timmer95}
{Timmer} J.,  {K{\"o}nig} M.,  1995, \aap, 300, 707

\bibitem[\protect\citeauthoryear{Titterington, Smith \& Makov}{Titterington
  et~al.}{1985}]{Titterington85}
Titterington D.~M.,  Smith A. F.~M.,    Makov U.~E.,  1985, Statistical
  Analysis of Finite Mixture Distributions.
Wiley, New York

\bibitem[\protect\citeauthoryear{{Trotta}}{{Trotta}}{2008}]{Trotta08}
{Trotta} R.,  2008, Contemporary Physics, 49, 71

\bibitem[\protect\citeauthoryear{{Uttley}, {McHardy} \& {Papadakis}}{{Uttley}
  et~al.}{2002}]{Uttley02}
{Uttley} P.,  {McHardy} I.~M.,    {Papadakis} I.~E.,  2002, \mnras, 332, 231

\bibitem[\protect\citeauthoryear{{Uttley}, {McHardy} \& {Vaughan}}{{Uttley}
  et~al.}{2005}]{Uttley05}
{Uttley} P.,  {McHardy} I.~M.,    {Vaughan} S.,  2005, \mnras, 359, 345

\bibitem[\protect\citeauthoryear{{van der Klis}}{{van der
  Klis}}{1989}]{vanderklis89}
{van der Klis} M.,  1989, in {{\"O}gelman} H.,  {van den Heuvel} E.~P.~J.,
  eds, Timing Neutron Stars {Fourier techniques in X-ray timing}.
p.~27

\bibitem[\protect\citeauthoryear{{Vaughan}}{{Vaughan}}{2005}]{Vaughan05}
{Vaughan} S.,  2005, \aap, 431, 391

\bibitem[\protect\citeauthoryear{{Vaughan}, {Edelson}, {Warwick} \&
  {Uttley}}{{Vaughan} et~al.}{2003}]{Vaughan03b}
{Vaughan} S.,  {Edelson} R.,  {Warwick} R.~S.,    {Uttley} P.,  2003, \mnras,
  345, 1271

\bibitem[\protect\citeauthoryear{{Vaughan} \& {Fabian}}{{Vaughan} \&
  {Fabian}}{2003}]{Vaughan03}
{Vaughan} S.,  {Fabian} A.~C.,  2003, \mnras, 341, 496

\bibitem[\protect\citeauthoryear{{Vaughan} \& {Uttley}}{{Vaughan} \&
  {Uttley}}{2005}]{Vaughan05b}
{Vaughan} S.,  {Uttley} P.,  2005, \mnras, 362, 235

\bibitem[\protect\citeauthoryear{{Vaughan} \& {Uttley}}{{Vaughan} \&
  {Uttley}}{2006}]{Vaughan06}
{Vaughan} S.,  {Uttley} P.,  2006, Advances in Space Research, 38, 1405

\bibitem[\protect\citeauthoryear{Venables \& Smith}{Venables \&
  Smith}{2009}]{r2}
Venables W.~N.,  Smith D.~M.,  2009, An Introduction to R.
http://cran.r-project.org/manuals.html

\bibitem[\protect\citeauthoryear{{V{\'e}ron-Cetty}, {V{\'e}ron} \& {Gon{\c
  c}alves}}{{V{\'e}ron-Cetty} et~al.}{2001}]{Veron01}
{V{\'e}ron-Cetty} M.-P.,  {V{\'e}ron} P.,    {Gon{\c c}alves} A.~C.,  2001,
  \aap, 372, 730

\bibitem[\protect\citeauthoryear{{Weedon}}{{Weedon}}{2003}]{Weedon03}
{Weedon} G.~P.,  2003, Time-Series Analysis and Cyclostratigraphy.
Cambridge University Press

\bibitem[\protect\citeauthoryear{{Whittle}}{{Whittle}}{1953}]{Whittle53}
{Whittle} P.,  1953, {Arkiv f\"{o}r Matematik}, 2, 423

\bibitem[\protect\citeauthoryear{Whittle}{Whittle}{1957}]{Whittle57}
Whittle P.,  1957, Journal of the Royal Statistical Society. Series B
  (Methodological), 19, 38

\end{thebibliography}

\appendix

%______________________________________________________________

\section{Simulating from the posterior}
\label{sect:mc}

Here we briefly discuss a method for simulating data from the posterior density,
which is useful for two main reasons.
For simple models with few parameters it may be possible to make inferences from the posterior without
the need for Monte Carlo simulations, e.g. by directly evaluating the posterior
density on a fine grid of parameter values. However, even in this case 
simulations from the posterior are needed in order to form the posterior
predictive distribution, and hence the distribution of a test statistic and its
posterior predictive $p$-value. For more complicated models or a greater number of 
parameters Monte Carlo methods may be necessary simply in order to calculate summaries of the posterior
(such as means and intervals). 

Markov chain Monte Carlo (MCMC) methods provide a powerful and popular method for
drawing random values from the posterior density. 
General introductions to
MCMC computations for Bayesian posterior calculations are given by \citet{Gelman04, Gregory05, Albert07}, 
and more thorough treatments may be found in \citet{Tierney94, Chib95, Gilks95, Gamerman97}.

The output of an MCMC calculation is a series of parameter values (or vectors) $\bt^t$ for 
$t=0,\ldots,L$ (where $L$ is the number of simulations performed, i.e. the length of the chain).
The Metropolis-Hastings MCMC algorithm generates a sequence of random draws as follows:
\begin{itemize}
\item
Draw a starting point in parameter space $\bt^0$ for which $p(\bt | \mathbf{I}, H) > 0$.

\item
Repeat for $t=1,\ldots,L$:

\begin{enumerate}

\item
Draw a proposed new parameter point $\bt^\ast$ from a \emph{proposal} distribution
$g(\bt^{\ast} | \bt^{t-1} )$ that is conditional only on the previous point $\bt^{t-1}$.

\item
Evaluate the ratio
\begin{equation}
 r = \frac{ p(\bt^{\ast} | \mathbf{I}, H) g(\bt^{\ast} | \bt^{t-1} ) }{
         p(\bt^{t-1} | \mathbf{I}, H)   g(\bt^{t-1} | \bt^{\ast} ) }
\end{equation}

\item
Set the new value of $\bt^t$
\begin{equation}
\bt^t = \left\{ \begin{array}{l l}
        \bt^{\ast} & \quad \mbox{with probability $\min(r,1)$} \\
        \bt^{t-1}  & \quad \mbox{otherwise} \\
        \end{array} \right.
\end{equation}

\end{enumerate}

\end{itemize}

In order to use this algorithm we need to have defined a proposal density function $g(\bt^{\ast}|\bt^{t-1})$, 
from which we can compute densities and draw random values, and 
be able to evaluate the posterior density at
any valid point in parameter space.
Notice that only the ratio of posterior densities need be calculated (to give $r$), meaning
that we can use the unnormalised posterior density $q(\bt | \mathbf{I}, H)$ in the computation.
The remarkable property of the MCMC algorithm is that the distribution of the output $\bt^t$ converges on the
target distribution $p(\bt | \mathbf{I}, H)$ for any form of proposal distribution \citep[see][for 
regularity conditions]{Tierney94,Gilks95}.
 
The choice of proposal density does however affect the speed of convergence to the target distribution 
(i.e. the efficiency of the MCMC calculation) -- the algorithm
will be most efficient when the choice of proposal density closely matches the posterior density.
We use as the proposal density a Normal random walk, specifically a
multivariate Normal distribution
centred on $\bt^{t-1}$ with the covariance $\Sigma$ from the Normal approximation to the posterior 
(see section \ref{sect:ba}). 
This is a  popular and well understood choice of proposal and has been discussed extensively in the
MCMC literature.
In fact, it is usually better to use $g(\bt^{\ast} | \bt^{t-1} ) = N(\bt^{t-1}, c\Sigma)$ where $c$ 
is a constant scale factor ($\gsim 1$) tuned to improve the efficiency of the calculation 
\citep[see section 11.9 of][]{Gelman04}.
In the present analysis we found $c=1.2$ to work well.
As the normal distribution is symmetric, i.e. $g(\bt^{i} | \bt^{j}) = g(\bt^{j} | \bt^{i})$,
the ratio $r$ simplifies to the ratio to the posterior densities $r = q(\bt^{\ast}|\mathbf{I},H)/q(\bt^{t-1}|\mathbf{I},H)$. 
(We also found that a multivariate Student's $t$-distribution worked comparably well, with a covariance
matrix $\Sigma \nu / (\nu -2)$, and $\nu=3$ degrees of freedom.)

One must take some care to ensure the output of the MCMC has reached its \emph{stationary} 
distribution and is efficiently generating draws from the complete posterior density. 
\citet{Gelman92}, \citet{Gilks95}, \citet{Kass98} and \citet{Gelman04} offer advice for checking the quality of the output. 
We calculate $J$ separate chains, each starting
from a different initial positions $\bt^0$ spread over the parameter space, and check for convergence before merging the results.
In order to remove any dependence on the initial position we retain only the second half of each chain.
We then compute the $\hat{R}$ statistic \citep{Gelman92, Gelman04}\footnote{In \citet{Gelman92} and the first edition of the book \citet{Gelman04} 
this statistic was called $\hat{R}^{1/2}$.}, which
compares the variance within each chain to the variance between the chains. With a sufficiently
large number of iterations $\hat{R}$ will be close to unity, which indicates that the chains have reached 
the desired stationary distribution. 

In practice we discarded the first half of each chain (the ``burn in'' phase) to remove any dependence on the starting position, and required $\hat{R} < 1.1$ 
from the remaining half of the chain before assuming the stationary 
distribution has been reached. We also inspect, for each chain, the acceptance rate, and the time series and histograms of the 
parameters, which may also reveal problems with convergence or efficiency of the chains.
Once convergence has been reached we combine the remaining $L/2$ points from each chain to yield $JL/2$ sets of parameter values
sampled from the posterior distribution.

Given a sufficient number of simulations, $\bt^t$, we can 
estimate the posterior distribution of any quantity of interest, such as the untransformed parameters, 
and from these estimate the posterior means (or modes, or medians) and credible intervals by calculating the
sample means and quantiles. If necessary we can also simulate
data $\mathbf{I}^{\rm rep}$ from the posterior predictive distribution by sampling parameters from the MCMC output, $\bt^{\rm rep}$, 
and for each point calculating $S(\bt^{\rm rep})$ and then randomly drawing periodogram 
points according to eqn \ref{eqn:pdist} (i.e. a scaled $\chi_2^2$ distribution).
We can then calculate the posterior distribution of any test statistic $T$ using these simulated data, and
hence calculate a $p$-value.

For the analysis of section \ref{sect:data} we performed an initial fit to the data, using a non-linear optimisation 
algorithm to find the posterior mode and covariance matrix, and used this to form the proposal distribution for the MCMC.
Five chains of length $30,000$ were generated from different locations randomly dispersed around the posterior mode.
The first half of each chain (the ``burn-in phase'') was discarded and,
after checking convergence was achieved (as measured using the diagnostics discussed above), the remaining $75,000$ values were merged
into a single sample.

%______________________________________________________________

\section{Simulation of time series}
\label{sect:sim-ts}

The calculation of a sample from the posterior, $p(\bt | \mathbf{I}^{\rm obs}, H)$, requires only the output from an
MCMC such as outlined above. Posterior predictive checks require the generation of simulated (replicated) periodogram data $\mathbf{I}^{\rm rep}$.
The simplest approach is to draw a random parameter vector from the posterior, $\bt^t$, generate the corresponding power
spectral density function $S_j(\bt^t)$ at frequencies $j=1,\ldots,N/2-1$ and multiply each of these by a random draw from 
the $\chi_2^2$ (exponential) distribution (equation \ref{eqn:pdist})
\begin{equation}
\label{eqn:prand}
  I_j^{\rm rep} = S_j(\bt^t) X_j /2,
\end{equation}
(where $X_j$ are independent random variables drawn from the $\chi_2^2$ distribution).
The periodogram at the Nyquist frequency $j=N/2$ should be
multiplied by a random draw from a $\chi_1^2$ distribution instead. (The zero frequency component can be safely given zero power.)

Time series may be generated by inverse Fourier transforming the randomised periodogram into the time domain (with time steps $\Delta T$), with appropriate phase
randomisation. However, the resulting Fourier transformed data are strictly periodic with a period $N$, and so there is a wrap-around effect 
where the start and end of the time series are forced to converge. Also, this procedure does not include any effects due to transfer of power
from frequencies just below or above the observed range. More realistic data may be generated 
from a posterior draw $\bt^t$ by calculating a power spectrum over a wider range of 
frequencies than are included in the data, e.g. over a frequency grid $f_k = k/V N \Delta T$ with $k=1,\ldots,K$, 
where $V \ge 1$ and $W \ge 1$ are the factors by which the lowest and highest frequencies are extended (respectively), 
and $K=VWN/2$.
The power spectral densities $S_k(\bt^t)$ may then be used in the algorithm of 
\citet{Davies87}\footnote{An equivalent algorithm for generating time series from a power spectrum was introduced to 
astronomy by \citet{Timmer95}.} to produce a time series. 

An alternative, but mathematically equivalent method, is 
as follows:
\begin{itemize}

\item
Generate $K-1$ random periodogram ordinates $I_k^{\rm rep}$ by multiplying the spectrum $S_k(\bt^t)$ with random $\chi_2^2$ variables as in 
equation \ref{eqn:prand}. (At the Nyquist frequency, $k=K$, use a $\chi_1^2$ variable instead of $X_k/2$.)

\item
Generate $K-1$ independent, random phases $\phi_k$ over the range $[-\pi/2, \pi/2)$ from a uniform distribution.
(At the zero and Nyquist frequency use $\phi_k=0$.)

\item
Produce a complex vector $F_k = A_k \exp(-i \phi_k)$ with arguments $A_k = \sqrt{I_k/2}$ and phases $\phi_k$.

\item
Extend the vector (of Fourier amplitudes and phases) to negative frequencies, 
setting the Fourier components for the negative frequencies 
$F_{-k} = F_k^{\ast}$ where the asterisk denotes complex conjugation.
(Note that the Fourier components are real valued at the zero and Nyquist frequencies.)

\item
Inverse Fourier transform the $\{ F_k \}$ from the frequency domain to the time domain.

\end{itemize}

The resulting series will be $2K (= VWN)$ points in length, with a
sampling rate of $\Delta T / W$ and a duration of $V N \Delta T$, and will have a mean of approximately zero. 
(Time series that more closely resemble those of accreting compact objects can be obtained using the exponential
transformation of \citealt{Uttley05}.)
One may then resample a segment of this to match the
sampling pattern of the observation, to give a time series of $N$ points, as required (and with no wrap-around effect). For most processes it should make little difference
whether the noise due to ``measurement error'' is included in the power spectrum, or excluded from the power spectrum 
and added at the resampling stage (e.g. by drawing the counts per bin from the Normal or Poisson distribution after appropriate normalisation
of the series). See \citet{Uttley02} for more on time series simulation.

\bsp

\label{lastpage}

\end{document}